\documentclass[letterpaper, 12pt]{article}

\usepackage[margin=1 in]{geometry}
\usepackage[doublespacing]{setspace}
\AtBeginEnvironment{tabular}{\doublespacing}
\AtBeginEnvironment{tabularx}{\doublespacing}

\usepackage{titlesec}
\titleformat*{\section}{\normalsize\bfseries}
\titleformat*{\subsection}{\normalsize\bfseries}
\usepackage{xcolor}

\usepackage{amssymb}
\usepackage{ulem}

\usepackage{graphicx}
\usepackage{subcaption}
\usepackage{multirow}
\usepackage{tabularx}
\usepackage{etoolbox}  
\usepackage{booktabs}
\usepackage[export]{adjustbox}

\usepackage{mathtools}
\numberwithin{equation}{section}
\usepackage[draft]{hyperref}

\newcommand{\strain}{\varepsilon}
\newcommand{\force}{\tilde{F}}
\newcommand{\peeq}{\ensuremath{\overline{\epsilon}} }

\newcommand{\bom}[1]{\boldsymbol{#1}} %
\newcommand{\sigb}{\bom{\mathrm{\sigma}}}

\newcommand{\vP}{\{\mathbf{P}\}}
\newcommand{\vF}{\{\mathbf{F}\}}
\newcommand{\vFs}[1]{\{\mathbf{F}_{#1}\}}
\newcommand{\vB}{[\mathbf{B}]}
\newcommand{\vBs}[1]{[\mathbf{B}_{#1}]}
\newcommand{\vBss}[2]{[\mathbf{B}_{#1}^{#2}]}
\newcommand{\vR}{\{\mathbf{R}\}}
\newcommand{\delR}{\{\bom{\Delta}_{\mathbf{R}}\}}
\newcommand{\delP}{\{\bom{\Delta}_{\mathbf{P}}\}}
\newcommand{\vG}{[\mathbf{G}]}

\begin{document}

\title{\onehalfspacing{Localized necking under global compression in two-scale metallic hierarchical solids}}

\author{
	Naresh Chockalingam S. \& Narayan K. Sundaram\footnote{Email: \url{nsundaram@iisc.ac.in}. Associate Professor of Civil Engineering}\\
	\small{Dept. of Civil Engineering, Indian Institute of Science, Bengaluru, India.}
	}
	
\date{}

\maketitle
\hrule

\begin{abstract}
Hierarchically structured cellular solids have attracted increasing attention for their superior mass-specific mechanical properties. Using a remeshing-based continuum finite element (FE) framework, we reveal that two-scale metallic hierarchical solids exhibit a distinct, localized deformation mode that involves necking and fracture of microscale tension members even at small global compressive strains (3--5\%). The tensile failure is always preceded by plastic buckling of a complementary compression member. This combined necking-buckling (NB) mode critically underlies the collapse of hexagon-triangle (HTH) hierarchical lattices over a wide range of relative densities and length-scale ratios and is also seen in diamond-triangle (DTH) lattices. In lattices with very slender microscale members, necking is prevented by a competing failure mode that involves coordinated buckling (CB) of multiple members. Our custom remeshing FE framework is critical to resolve the localized large plastic strains, ductile failure, and complex local modes of deformation (including cusp formation) that are characteristic of the NB mode. A theoretical buckling analysis supports the inevitability of the NB and CB modes in HTH lattices. The occurrence of the NB mode has consequences for energy absorption by two-scale hierarchical solids, and hence influences their design.
\end{abstract}
\clearpage
\section{Introduction}
Natural load-bearing solids as diverse as bones~\cite{lakes_1993}, tendons~\cite{fratzl_2007}, wood~\cite{fratzl_2007}, nacre~\cite{barthelat_2007}, and sea-sponge skeletons~\cite{aizenberg_2005} exhibit remarkably high specific stiffness, strength, and toughness on account of their multiscale hierarchical architecture. These superior properties, in turn,  have inspired engineered hierarchical solids that similarly exhibit enhanced stiffness and strength~\cite{zheng_2016}, resilience~\cite{meza_2015}, defect-tolerance~\cite{sen_2011}, and energy absorption~\cite{chen_2018} in comparison to their non-hierarchical counterparts. 

Two-scale hierarchical cellular solids (THCS) are an important subclass of such engineered solids; these are best described as honeycomb-like cellular structures with an additional level of substructure in their cell walls~\cite{banerjee_2014}. In particular, THCS with bending-dominated macro-cells like hexagons paired with stretching-dominated micro-cells like triangles offer a combination of high stiffness and toughness besides much higher specific energy absorption~\cite{chen_2018}. This makes them better-suited for energy absorption applications than single-scale cellular solids as they overcome some of the well-known limitations of the latter. For instance, bending-dominated cellular solids have low stiffness~\cite{fleck_2010} and stretching-dominated cellular solids show a high initial peak force~\cite{deshpande_2001, gibson2003cellular}. 

While the in-plane elastic properties of various THCS are well-established by theoretical analyses~\cite{taylor_2011, ajdari_2012, sun_2014, banerjee_2014, haghpanah_2013}, finite-element (FE) simulations are indispensable to study their inelastic and large-deformation response. In these cases, hierarchical solids are typically modelled using beam~\cite{tao_2019} and shell~\cite{qiao_2016, yin_2018, liu_2022} finite elements. In particular, the collapse load of hierarchical lattices with hexagonal macrostructure and triangular substructures (referred to henceforth as HTH) was found to be higher than both single-scale hexagons and triangles~\cite{qiao_2016}. Subsequent simulations~\cite{yin_2018, tan_2019, zhan_2022} reinforced the superiority of triangular substructures. Shell FE simulations have also been used to study the energy absorption characteristics of lattices with diverse macrostructures including kagome~\cite{wang_2021}, re-entrant~\cite{zhan_2022}, aux-hex~\cite{xu_2022}, and graded HTH~\cite{liu_2022}. In most of these studies, the substructure was triangular. 

The aforementioned studies mainly focus on the global compressive response (e.g. plateau stress).  However, the presence of substructures gives rise to deformation modes that can involve highly-localized strain and failure. This has been reported in experiments on self-similar polymeric square honeycombs~\cite{tao_2019}, as well as in unit-cell simulations~\cite{chen_2016} of two-level metallic hexagonal / kagome (HKH) solids. In particular, a necking-like mode was observed in the latter~\cite{chen_2016}, which strongly suggests the potential for member failure. Moreover, Liu et al. \cite{liu_2022} report the occurrence of member fracture at early stages in uniaxial compression experiments on HTH, and also note the inability of shell FE simulations to capture the observed fracture. 

Strain localization by necking and subsequent tensile failure would critically affect the load-carrying capacity of macroscopic members as well as the global ductility, and must therefore be considered during substructure design. Importantly, capturing such a failure mode demands a full-scale continuum FE analysis. While there have been several continuum FE studies of hierarchical solids, these have either (1) focused on analysing single unit cells with periodic boundary conditions~\cite{chen_2016, chen_2018} or (2) are multi unit-cell simulations that focus on global crashworthiness~\cite{li_2020, zhan_2022, usta_2023} without modelling local member failure and post-failure deformation. 

\noindent
The current paper uses high-fidelity continuum FE simulations with remeshing to resolve, up to the smallest length-scale, the complex local modes of deformation and failure encountered in uniaxial compression of metallic hierarchical solids. It will be seen that remeshing is crucial to mitigate the mesh distortion due to the large localized plastic strains, member failure, inter-wall contacts, and self-contacts in these simulations. Similar remeshing schemes have been successful in mitigating severe element distortion in other areas of large-strain plasticity including metal forming~\cite{cheng_1986}, machining~\cite{vandana_sundaram_2020}, and deep indentation~\cite{das_sundaram_2024}. 
Importantly, we uncover a combined necking-buckling (NB) mode of deformation in HTH lattices and investigate its prevalence, mechanics, and consequences. Notably, a theoretical analysis supports the existence of such a mode in HTH lattices over a wide parametric range.
\section{Methods}	
Finite element (FE) simulations of uniaxial compression of representative two-scale hierarchical solids  are carried out using the Abaqus explicit dynamics solver~\cite{abaqus} in double precision mode using a custom remeshing framework as described below. 
\subsection{Remeshing scheme}
\begin{figure}
	\centering
	\includegraphics[width=0.65\textwidth]{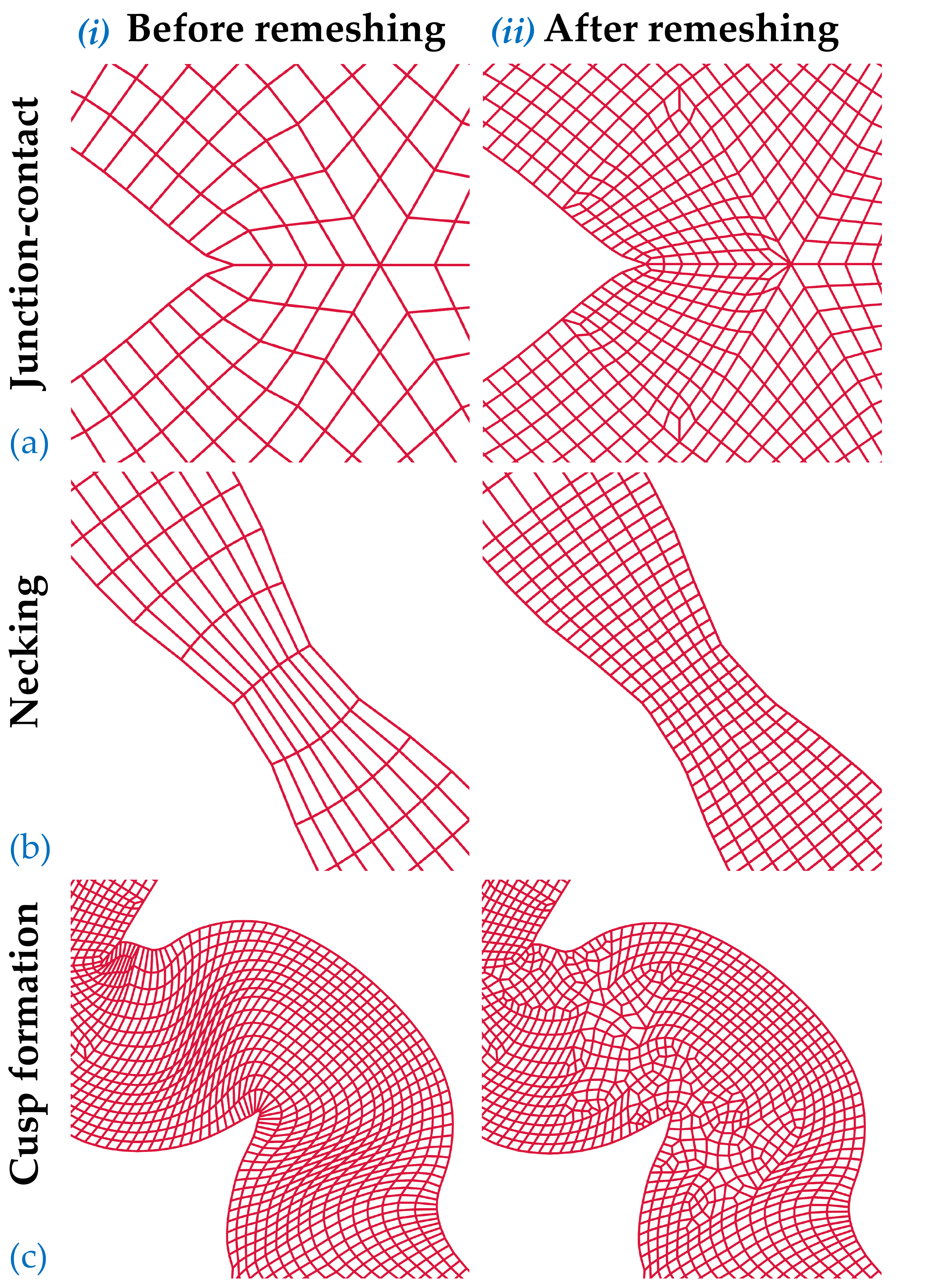}
	\caption{FE meshes \textit{(i)}-before and \textit{(ii)}-after remeshing for (a) inter-member contact emanating at the junction between two members, (b) necking of a tension member, (c) self-contact and cusp formation in a stocky microscale member undergoing large flexure. Remeshing is essential in all three cases.}
	\label{fig::remeshing}
\end{figure}

Plane strain, continuum finite elements are used to discretise the two-scale hierarchical solids (THCS) as they can accurately model junctions and contacts, and are applicable regardless of member slenderness. However, it will be seen that plastic deformation in THCS has distinctive features like contact between microscale members that emanates at a junction (Fig.~\ref{fig::remeshing}(a)), necking (Fig.~\ref{fig::remeshing}(b)), and cusp formation (Fig.~\ref{fig::remeshing}(c)). As a consequence, any initial mesh is eventually severely distorted and becomes unusable.  

For example, the distorted meshes in the left column (\textit{i}) of cases (a-c) in Fig.~\ref{fig::remeshing} show multiple defects, including quadrilaterals which have degenerated into triangles, elements with high aspect ratios, and highly-skewed elements. Rediscretising these regions with a new mesh as shown in column (\textit{ii}) is critical to continue the simulations. In addition, remeshing can optimize the computational cost if one incorporates adaptive refinement of critical zones. To this end, a custom-made, Python-based remeshing framework is proposed in the current study which encompasses (a) an automatic, all-quad mesh generator, (b) a mesh-to-mesh transfer module, and (c) an automatic contact surface detector. The remeshing algorithm is as follows:
\begin{enumerate}
	\item Identify the region $\mathcal{R}$ to be remeshed as a set of (possibly distorted) elements.
	\item Delineate the boundary $\partial \mathcal{R}$ of $\mathcal{R}$ which consists of exterior segments $\partial\mathcal{R}^e$ and segments $\partial\mathcal{R}^i$ shared by the neighbouring non-remeshed regions.
	\item Parametrise $\partial\mathcal{R}^e$ using linear interpolants, and rediscretise with the desired mesh size.
	\item With the redefined $\partial\mathcal{R}^e$ and the undisturbed $\partial\mathcal{R}^i$ bounding the region $\mathcal{R}$, generate a new mesh in the interior of $\mathcal{R}$ (discussed in section~\ref{subsec:meshing}).
	\item Transfer the nodal and integration point (IP) variables from the old mesh onto the new mesh (discussed in section~\ref{subsec:transfer}).
	\item Redefine all potential contact surfaces using a suitable boundary-detection algorithm and resume the simulation.
\end{enumerate}
The kinematics of the deformation modes shown in Fig.~\ref{fig::remeshing} are clearly beyond any structural theory. Furthermore, features like cusp formation demand topological changes in the mesh and cannot be handled by mere geometric regularization~\cite{das_sundaram_2024}. 
\subsection{Solution transfer operator}
\label{subsec:transfer}
A transfer operator in a remeshing scheme is a rule to derive the nodal and integration point (IP) variables in the new mesh, given the old mesh and its corresponding solution variables. A wide range of transfer operators are available, including\footnote{in roughly increasing order of complexity} weighted-averaging~\cite{shepard_1968, habraken_1990, das_sundaram_2024}, parametric inversion and interpolation~\cite{crawford_1989, lee_1994, peric_1996},~projection~\cite{hinton_1974, petersen_1997, leger_2014}, super-convergent patch recovery~\cite{zienkiewicz_1992, boroomand_zienkiewicz_1999, kumar_2015}, and variationally consistent transfer~\cite{ortiz_1991}. The current study adopts a local projection scheme that is customized for reduced-integrated quad-element  meshes. The aim is to preserve extreme values (minimal diffusion) and to not rely on the FE interpolation functions of the (generally-distorted) old mesh.

The new boundary nodes are constrained to lie along the edges of the old mesh as their positions and velocities are derived by linear interpolation along $\partial \mathcal{R}^{e}$. 
In the interior, mesh-to-mesh projection of the nodal fields $\left\{v_{i}(x,y)\right\}$ and the IP variables $\left\{q_{j}(x,y)\right\}$ is carried out by minimizing the squared error $\mathcal{E}$, defined as:
\begin{equation}
	\mathcal{E} = \sum\limits_{i}^{\substack{\text{nodal}\\ \text{variables}}}\int\limits_{\mathcal{R}}\left(v_{i}^{\text{new}} - v_{i}^{\text{old}}\right)^{2}dA + 
	\sum\limits_{j}^{\substack{\text{IP} \\ \text{variables}}}\int\limits_{\mathcal{R}}\left(q_{j}^{\text{new}} - q_{j}^{\text{old}}\right)^{2}dA							
\end{equation}
subject to the constraint $ v_{i}^{\text{new}}\left(x_{c}, y_{c}\right) = {v}_{i}^{\text{old}}\left(x_{c}, y_{c}\right) $ for all nodal points $\left(x_{c}, y_{c}\right)$ on the boundary $\partial\mathcal{R}$. 
Consistent with the assumptions of reduced integration, $q_j^{\text{old}}$ and $q_j^{\text{new}}$ are interpolated in an element-wise constant manner (as in~\cite{rashid_2002}), and the FE shape functions of the new isoparametric four-noded elements are used to interpolate $v_i^{\text{new}}$. As the old mesh is distorted, $v_i^{\text{old}}$ is interpolated using multiquadrics~\cite{hardy_1971}, a popular method for scattered-data interpolation, instead of the old FE shape functions. Tests using synthetic functions show that the numerical diffusion in the current transfer scheme is minimal if the new mesh is finer than the old mesh.
\subsection{Lattice geometries, loading, and boundary conditions}
\label{subsec:bcs}
\begin{figure}
	\centering
	\includegraphics[width = \textwidth]{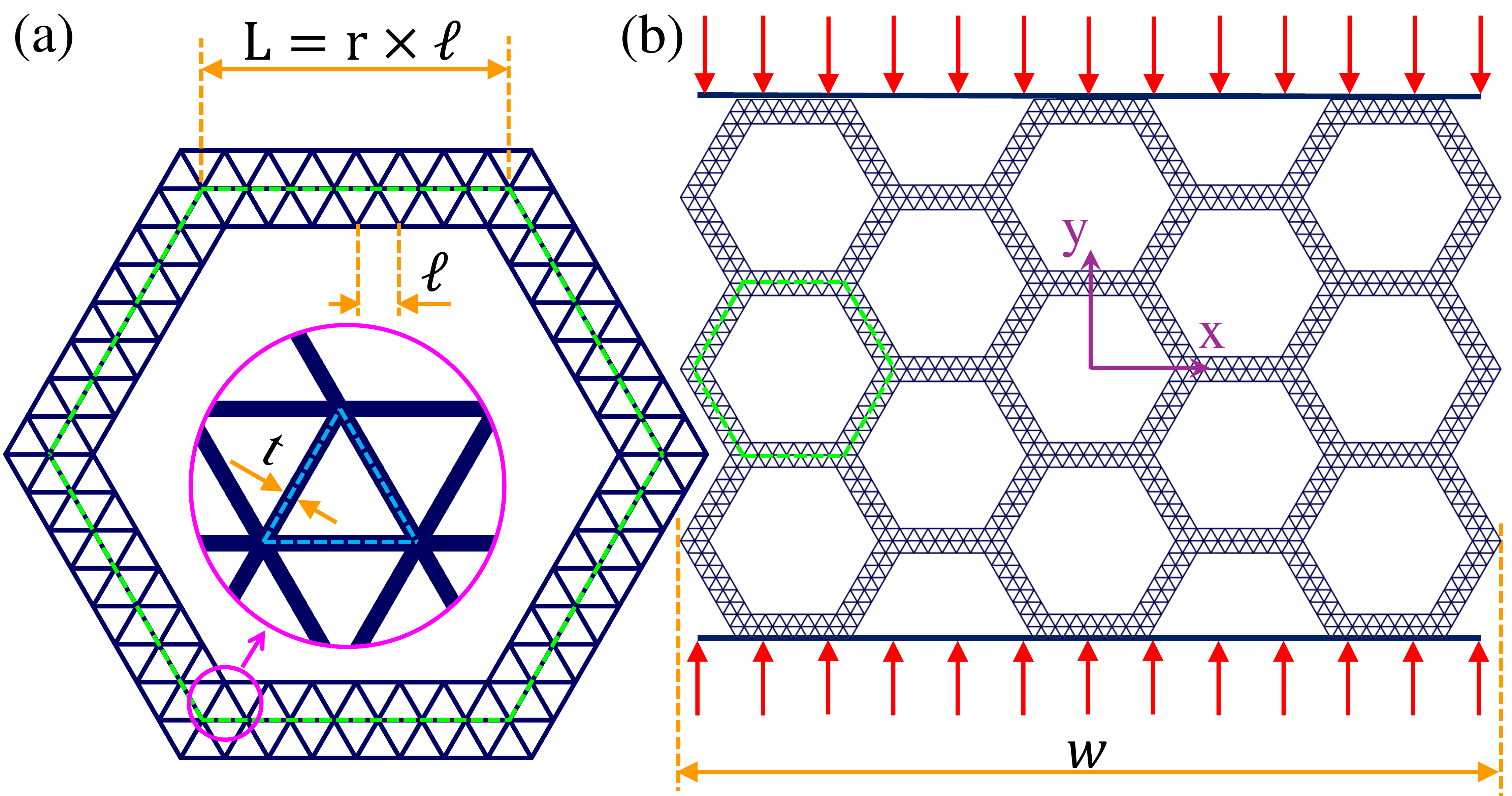}
	\caption{(a) A unit cell of a hexagon-triangle hierarchical (HTH) lattice (b) Schematic of a multi-cell HTH specimen of width $w$ subjected to uniaxial compression along the y-axis}
	\label{fig::schematic}
\end{figure}

Metallic hexagon-triangle hierarchical (HTH) lattices and diamond-triangle hierarchical lattices (DTH) are considered in the present study. A constituent unit cell of HTH, and the tessellation of cells that define the lattice are shown in Figs.~\ref{fig::schematic}(a) and~\ref{fig::schematic}(b) respectively. The key  geometric parameters of the unit cell in Fig.~\ref{fig::schematic}(a) are the thickness $t$ of microscale members, the side $\ell$ of the triangular substructures, the hierarchical length-scale ratio $r$, which is the number of substructures along the length of macroscale cell wall, and $N$, the number of substructure units along its thickness. $N$ = 2 for all the simulations in this paper. The nominal slenderness $\ell/t$ of the microscale members is denoted by $\lambda$. The relative density $\rho$ of HTH is given in terms of the parameters $r$, $N$, and $\lambda$ as: 
\begin{equation*}
	\rho = 1 - \left(1 - \dfrac{N}{2r} - \dfrac{1}{\sqrt{3}r\lambda}\right)^2 - \left(\dfrac{N}{r} - \dfrac{N^2}{4r^2}\right)\left(1 - \dfrac{\sqrt{3}}{\lambda}\right)^2
\end{equation*}
The DTH lattice has diamond macro-cells with triangular substructures. (See Fig.~\ref{fig::dth}(c) for  the DTH geometry, with the inset showing a zoomed-in view of a junction). Its geometry is characterized by the same four parameters as HTH, viz., $t$, $\ell$, $r$, and $N$. However, since the substructures are right-angled isosceles triangles in DTH (unlike the equilateral triangles in HTH), $\ell$ is the length of the hypotenuse of these triangles. 
\noindent
As shown in Fig. \ref{fig::schematic}(b), a multi-cell specimen of HTH is subjected to uniaxial compression. Symmetry boundary conditions are enforced along the y-axis to model the right half ($x \geq 0$) of the solid. In order to confirm that the observed micro-mechanical modes are unaffected by the assumed global symmetry, full-scale simulations without the symmetry assumption are also carried out (see Sec.~\ref{subsec::robustness}). The applied global strain rate is kept at 0.01 s$^{-1}$ in all simulations to simulate quasi-static conditions. Mass-scaling is employed to increase the stable time increment while still ensuring that the kinetic energy is negligible throughout the simulation (see \ref{app::quality} for a discussion). 

The contact boundary conditions between the platens and the specimen are enforced by a master-slave contact-pair algorithm. The interface is treated as nearly frictionless (with Coulomb coefficient $\mu = 0.01$). The definitions of potential inter-wall and intra-wall contacts are allowed to evolve with the deformation. These interior contacts used a balanced contact-pair formulation and have $\mu = 0.10$. All contact boundary conditions are kinematically enforced for accuracy.  
\subsection{Deformation-driven mesh generation}
\label{subsec:meshing}
Plane strain, reduced integrated, four-noded, continuum elements with hourglass control (CPE4R) are used to discretise the solid. 
The initial mesh, with 8 elements though the thickness $t$ of a microscale member, is sufficiently fine to capture moderate-curvature elastoplastic flexure. This number is increased to as many as 16 elements (as in Fig.~\ref{fig::remeshing}c-\textit{i}) through the thickness as the member curvature increases. The initial mesh is generated using the mapped-element technique~\cite{haber_1981} after partitioning the lattice geometry into convex quadrilaterals\footnote{This technique projects the mesh from a parent bi-unit square onto these quadrilaterals by means of bilinear  interpolation functions.}. 

The sides of these quadrilaterals eventually become curved as the solid deforms. Provided the curvatures of these sides remain low and the internal angles are not too obtuse (or reflex), we map a structured mesh from a parent bi-unit square to the curved quadrilateral using a bilinearly-blended projector~\cite{haber_1981, lee_1994}. However, once the curvature becomes large or when concavities develop, such a mapping is not possible with satisfactory element quality. At this stage, we use an unstructured, advancing front algorithm~\cite{lee_1994_comput_struct} for mesh generation. Here an all-quad mesh is obtained in two-steps -- a constrained Delaunay triangulation using the \textit{triangle} package~\cite{shewchuk_1996}, followed by systematic recombination of triangles.

\subsection{Constitutive relations and ductile damage models}
A representative, moderately ductile metal (aluminium alloy AA5052-H32), with Young's modulus $E = $ 69 GPa and Poisson's ratio $\nu =$ 0.33 is considered for all simulations in the current study. Assuming von Mises-associated flow with isotropic hardening, the yield stress $\bar{\sigma}$ is given by a modified Voce equation~\cite{lloyd_1982} as a function of the effective plastic strain $\bar{\epsilon}$ and strain rate $\dot{\bar{\epsilon}}$:
\begin{equation}
	\bar{\sigma}\left(\bar{\epsilon}, \dot{\bar{\epsilon}}\right) = \left[\sigma_{s} - \left(\sigma_{s}-\sigma_{0}\right)\exp\left(-N\left(\bar{\epsilon} + \bar{\epsilon}_{0}\right)^p\right)\right]\beta(\dot{\bar{\epsilon}})
	\label{eq::flow}
\end{equation}
The rate-hardening factor $\beta(\dot{\bar{\epsilon}}) $ and the values of the parameters appearing in Eq.~\eqref{eq::flow} are given in \ref{app::material}.
A simple ductile failure criterion is incorporated to model localized fracture in these solids. Cracks are approximated by progressive deletion of elements that attain a critical fracture strain, i.e. $\bar{\epsilon} = \bar{\epsilon}_f$. The fracture strain is a function of the stress triaxiality $\eta = p /\bar{\sigma}$ as given\footnote{Here $p$ is the hydrostatic stress $p$ and $\bar{\sigma}$ the von Mises equivalent stress. } in Table~\ref{tab::damage} in \ref{app::material}. 
\section{FE Simulation Results}
\label{sec:results}
The remeshing-based continuum FE framework thus described is deployed to study the uniaxial compression response of HTH lattices over a range of relative densities $\rho$ and scale ratios $r$. Both the global (macroscale) response and their underlying local (microscale) mechanisms are revealed by these simulations.
\subsection{Global response of a representative HTH lattice}
\label{subsec:global}
\begin{figure}[h]
	\centering
	\begin{subfigure}{0.45\textwidth}		
		\includegraphics[width = \textwidth, left]{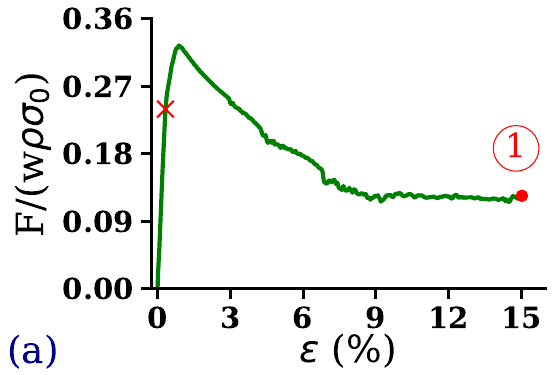}
	\end{subfigure}
	\begin{subfigure}{0.54\textwidth}
		\includegraphics[width = \textwidth, right]{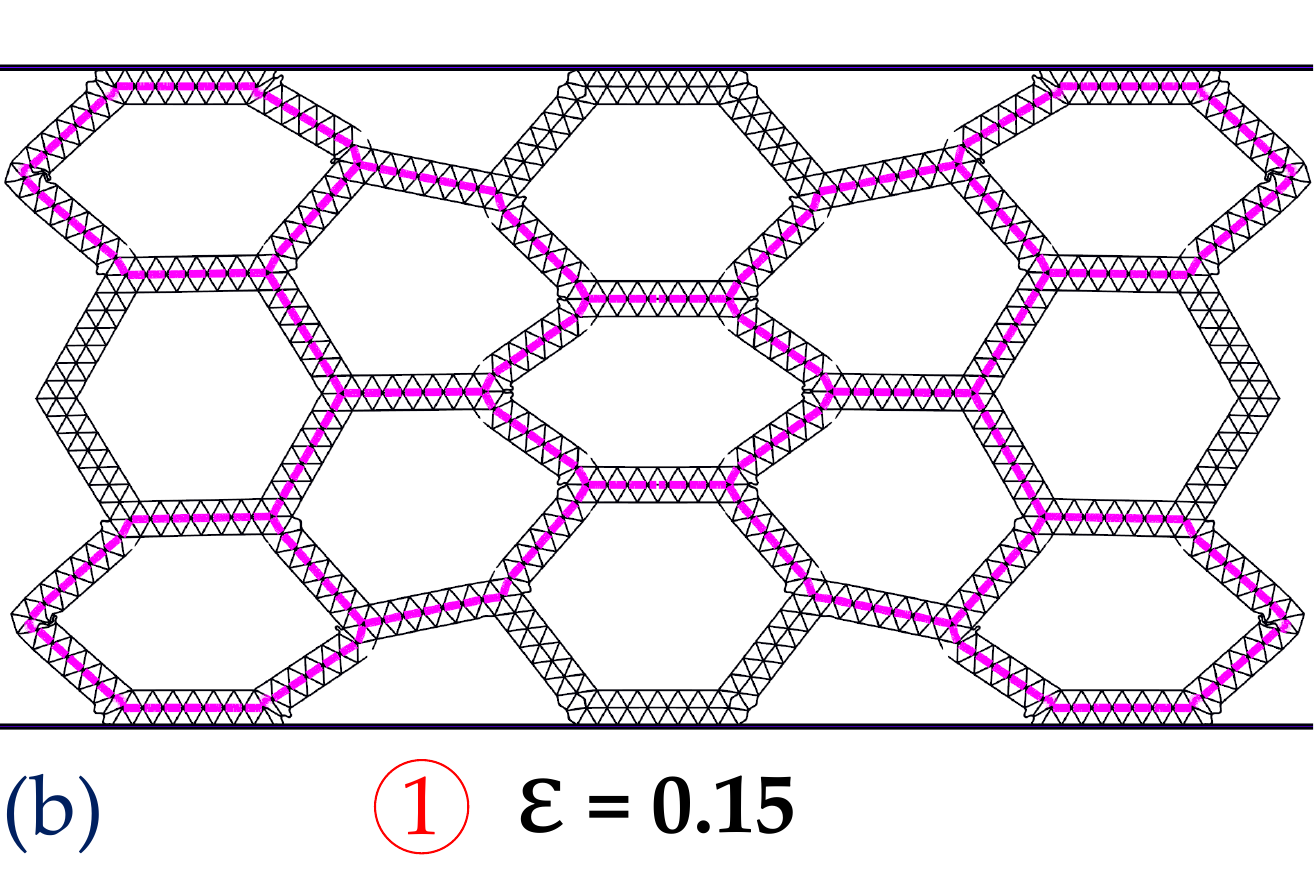}
	\end{subfigure}
	\caption{(a) Global force-strain response of an $r=7$, $\rho=0.1$ HTH lattice subjected to uniaxial compression (b) Deformed geometry of the HTH lattice at a global strain of $\strain = 0.15$; the centrelines of plastically deformed macro-cell walls are highlighted in magenta.} 
	\label{fig::mech1_global}
\end{figure}
\noindent
Fig.~\ref{fig::mech1_global}(a) shows the global uniaxial compression response of an HTH lattice with $\rho$ = 0.1 and $r=7$. Here, $\force = {F}/{w\rho\sigma_{0}}$ is a non-dimensionalised force and $\strain$ is the percentage nominal global strain. The initial response is linearly elastic, with a steep rise in force up to a strain of only $\strain = 0.3$\% (indicated by the $\times$). There is a subsequent and visible reduction in stiffness  (by about 75\%) due to plastic yielding of critical compression members. Upon further loading, a limit (peak) force $\force_{\text{p}}$ of 0.324 is attained at $\strain = 0.9$\%. This is followed by a steady post-peak softening that stabilizes to a constant plateau force of about 0.122 by a strain of $\strain = 10$\%. 

Fig.~\ref{fig::mech1_global}(b) shows the deformed geometry of the HTH lattice at a global strain $\strain$ of 0.15. The deformation pattern is non-uniform as the highlighted cells (in magenta) accommodate most of the imposed strain, with several other unit cells remaining relatively undeformed. Within the highlighted cells, the macro-cell walls have rotated almost rigidly as their ends undergo severe deformation. The precise nature of the near-junction deformation is not evident at the length-scale of Fig.~\ref{fig::mech1_global}(b), and requires one to zoom-in to the unit-cell scale.

\subsection{Localized necking under global compression and the necking-buckling (NB) mode}
\label{subsec:NB}
\begin{figure}[ht]
	\centering
	\includegraphics[width=\textwidth]{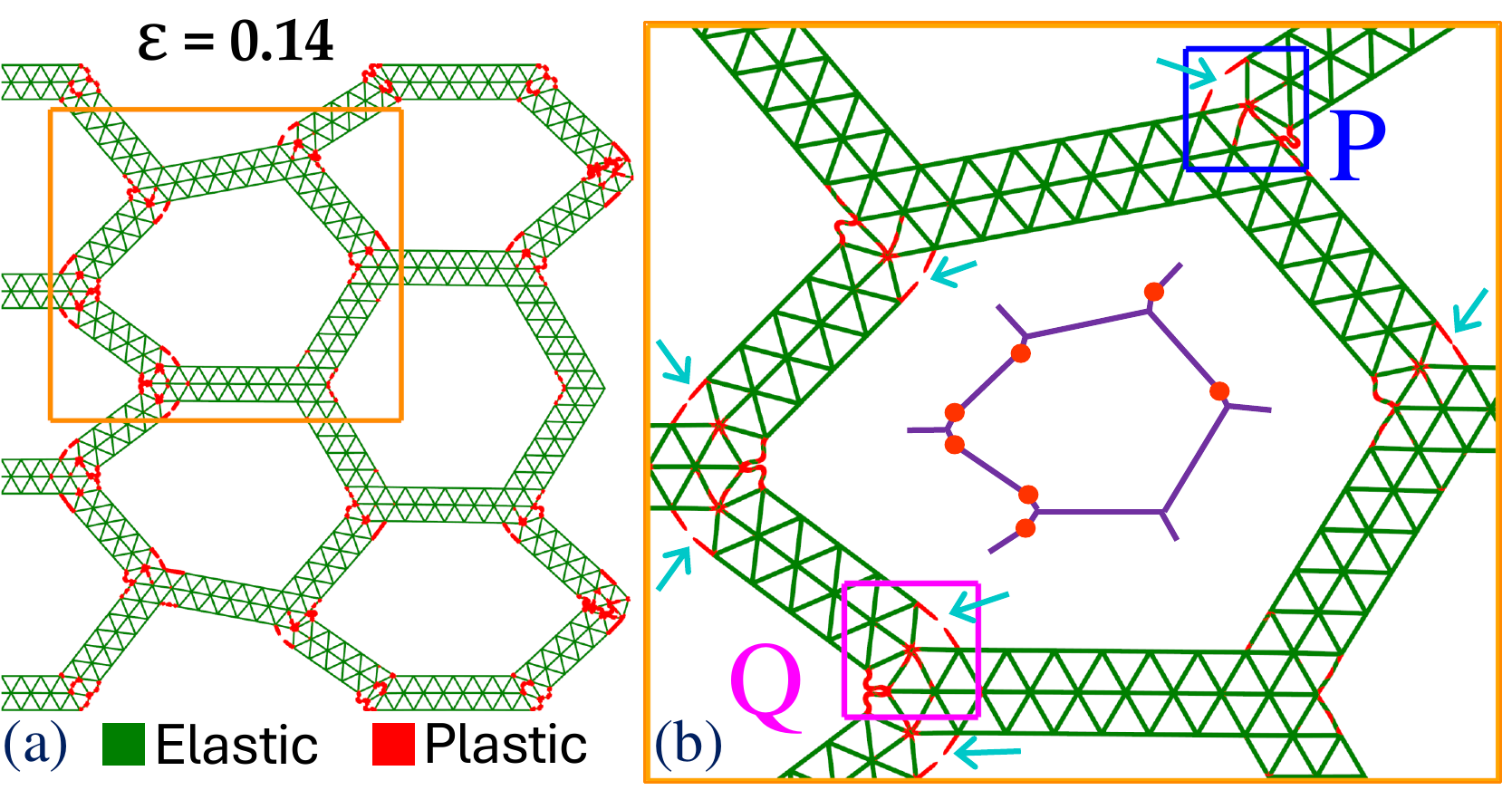}
	\caption{(a) Deformed configuration of a $\rho=0.1$, $r=7$ HTH lattice at a global strain of $\strain=0.14$. The elastic (green) and plastic (red) members are shown, with the latter highlighted for visibility (b) Zoomed-in view (to scale) of the deformed unit cell inside the box in panel (a).}
	\label{fig::mech1_elastic_plastic}
\end{figure}

Figs.~\ref{fig::mech1_elastic_plastic}(a) and~\ref{fig::mech1_elastic_plastic}(b) show successively zoomed-in views of some of the deformed macro-cells, with the microscale members colour-coded to show the elastic (green) and plastic (red) zones at a global $\strain$ of 0.14. A cut-off plastic strain value of $\peeq = 0.01$ is used to demarcate the plastic zones. Evidently, only a small fraction (about 6\%) of the members have yielded, with a large majority of them remaining elastic. Further, the yielded members are found exclusively near the ends of the macro-cell walls. These highly localized, near-junction plastic zones are responsible for the almost rigid rotation of macro-cell walls described earlier, and are reminiscent of plastic hinges in framed structures. 
A closer examination of these zones (Fig.~\ref{fig::mech1_elastic_plastic}b) reveals an interesting phenomenon : fractured members (indicated by the cyan arrows) under global compression. Localized fracture under global compression has not previously been simulated in two-scale hierarchical solids. We therefore study the time evolution of the deformation of small-scale members in two such critical zones (labelled `P' and `Q' in Fig.~\ref{fig::mech1_elastic_plastic}b) in order to uncover the mechanisms leading to fracture. 
\begin{figure}
	\centering	
	\includegraphics[width=\textwidth]{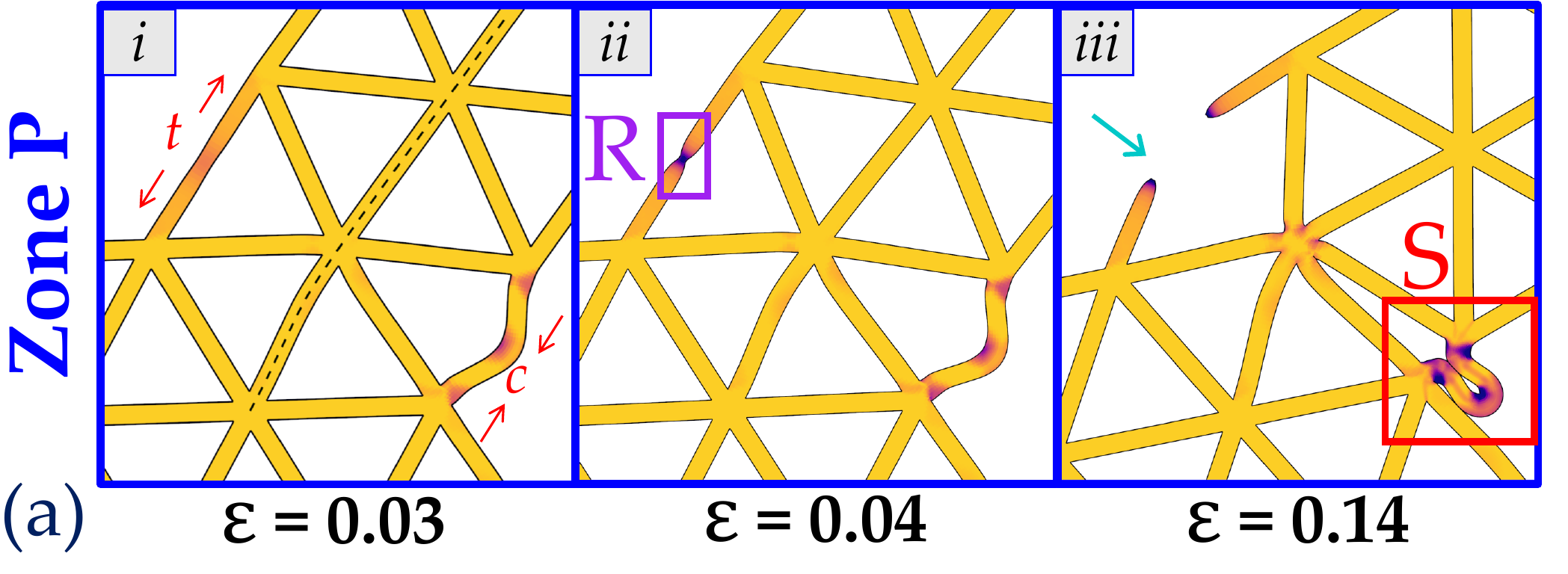}
	\includegraphics[width=\textwidth]{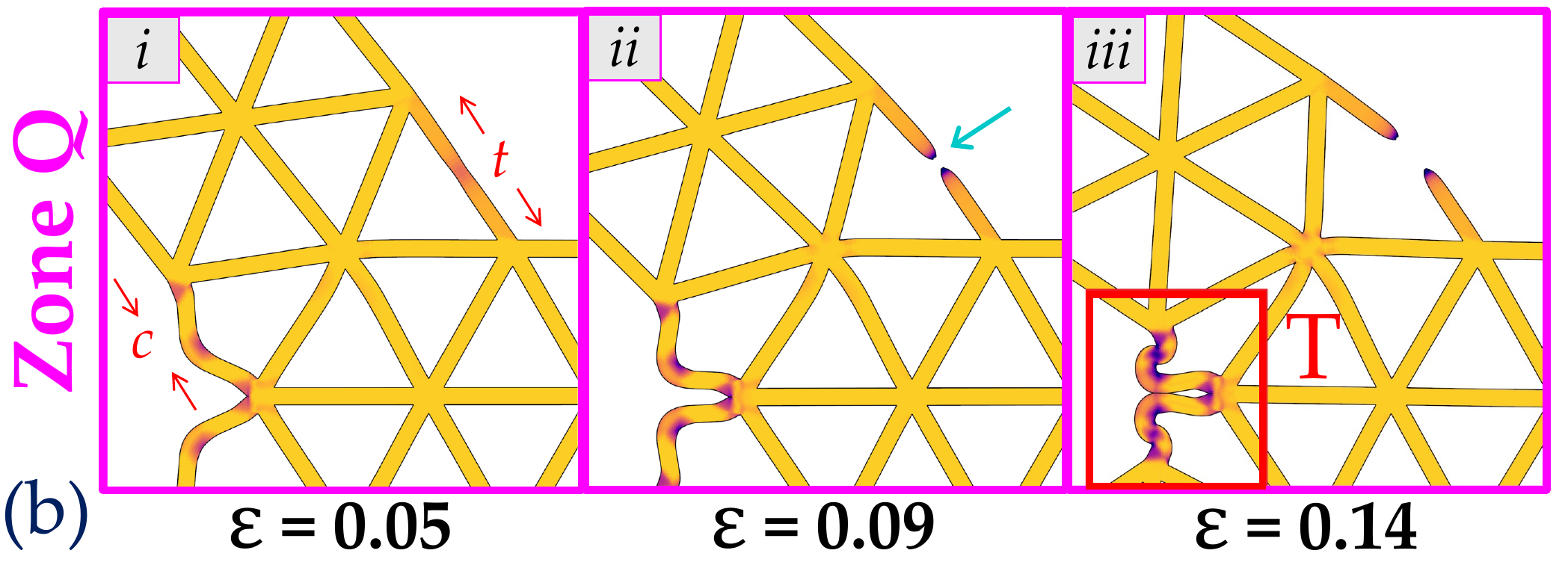}
	\includegraphics[width=\textwidth]{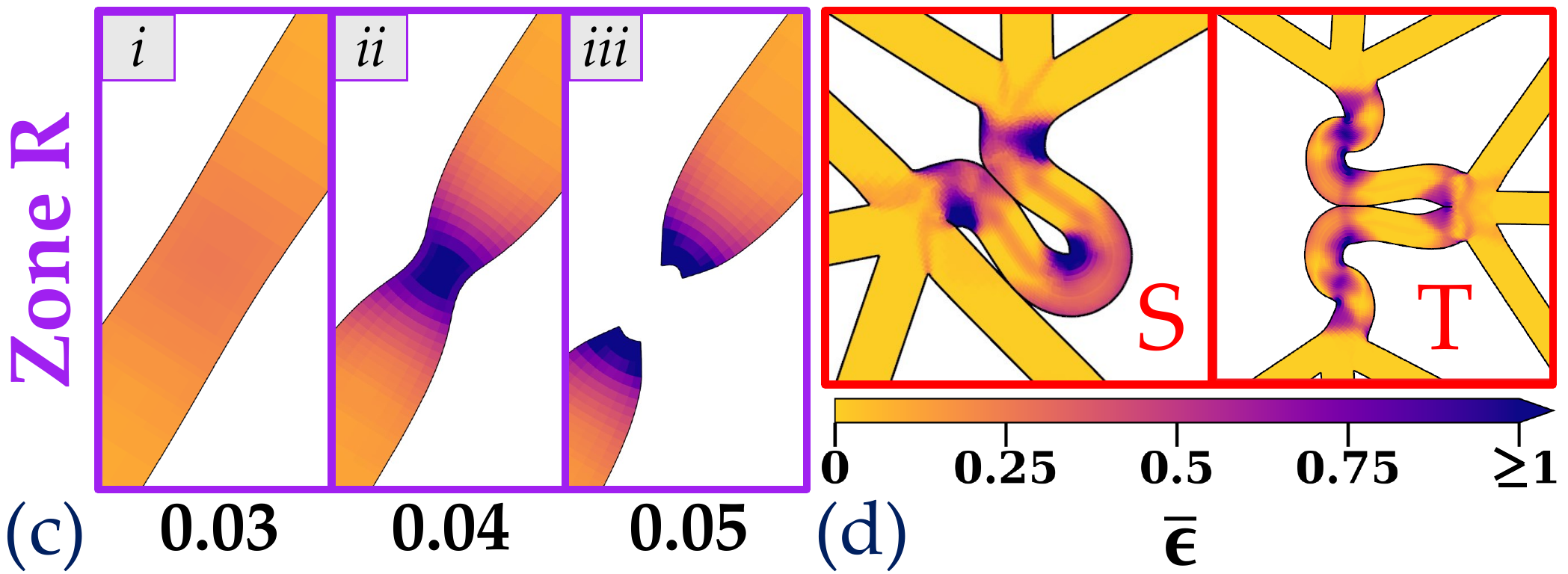}
	\caption{The necking-buckling (NB) mode in an $r=7$, $\rho=0.1$ HTH lattice. Effective plastic strain field $\peeq$ is superimposed on the deformed geometry in all sub-panels. 
		(a) Frames (a-\textit{i})--(a-\textit{iii}) show the $\peeq$ evolution in the near-junction zone `P' in Fig.~\ref{fig::mech1_elastic_plastic}(b) as global strain $\strain$ increases from 0.03 to 0.14. 
		Note the necking of member `$t$' in (a-\textit{ii}) and its fracture in (a-\textit{iii}). 
		(b) Frames (b-\textit{i})--(b-\textit{iii}) similarly show the NB mechanism in zone `Q' in Fig.~\ref{fig::mech1_elastic_plastic}(b) as $\strain$ increases from 0.05 to 0.14.
		(c) Frames (c-\textit{i})--(c-\textit{iii}) show the initiation, growth, and fracture of the neck in the boxed zone `R' in (a-\textit{ii}) as $\strain$ increases from 0.03 to 0.05. 
		(d) Zoomed-in views of zones `S' and `T' in panels (a-\textit{iii}) and (b-\textit{iii}) respectively, showing self-contact and inter-member contact.}		
	\label{fig::mech1_local}
\end{figure}

\noindent
Fig.~\ref{fig::mech1_local}(a) presents the deformation history of members in zone `P' in the sequence of frames (a-\textit{i})--(a-\textit{iii}), with the global strain increasing from 0.03 to 0.14. The superimposed background colour depicts the effective plastic strain field, $\peeq$.
As seen in Fig.~\ref{fig::mech1_local}(a-\textit{i}), the two extreme members indicated by `$t$' and `$c$' on opposite sides of the centreline (dashed line) of the rotating macro-cell wall have yielded at a small global strain of $\strain=0.03$. The members `$t$' and `$c$' act as chord members of a truss subjected to flexure, and are evidently under tension and compression respectively. Further, the member `$c$' has buckled plastically, with plastic deformation concentrated near its ends and mid-span; the maximum value of $\peeq$ is 0.5. 
The tension member `$t$' also yields, and quickly develops an inhomogeneous $\peeq$ distribution with a maximum value of $\peeq = 0.24$ attained at its mid-span. This is accompanied by a slight but distinct reduction in cross-section, thereby signifying the onset of necking (localization). With further loading ($\strain=0.04$) the strain field in the member `$t$' localizes further, sharply increasing the curvature of the neck as shown in Fig. \ref{fig::mech1_local}(a-\textit{ii}). Simultaneously, the compression member `$c$' undergoes further plastic buckling, with a $\max{(\peeq)}$ of 0.68. The tension member `$t$' cannot sustain indefinite strain localization, and eventually undergoes ductile fracture and member separation as seen in Fig.~\ref{fig::mech1_local}(a-\textit{iii}). It is also seen that the macro-cell walls continue to rotate about the plastic zones with reduced resistance, with the buckled member `$c$' folding smoothly in upon itself. 

As seen in Fig.~\ref{fig::mech1_local}(b), the deformation history of members in zone `Q' is very similar to that of zone `P'. The necking and buckling members are swapped in zone `Q' as the associated macro-cell wall rotates counter-clockwise, unlike the clockwise rotation in zone `P'. 
Besides `P' and `Q', a majority of the plastic zones of the HTH lattice exhibit this necking-buckling (NB) mode. In all cases, the rotation of macro-cell walls is enabled by plastic buckling of a compression member coupled with the necking of a corresponding tension member. 

It is instructive to examine the evolution of the neck more closely. Fig.~\ref{fig::mech1_local}(c) zooms in on the spatial zone labelled `R' in Fig.~\ref{fig::mech1_local}(a-\textit{ii}). The sequence of frames \textit{(i-iii)} in Fig.~\ref{fig::mech1_local}(c) have $\strain$ = 0.03, 0.04, and 0.05 respectively. The superimposed colour again shows $\peeq$. As the neck grows, the sides of member `$t$' develop rapidly increasing curvature, and the distribution of $\peeq$ through the thickness becomes non-uniform with its highest value ($\peeq = 1.2$) at the centre of the neck, as shown in Fig.~\ref{fig::mech1_local}(c-\textit{ii}). As a result, fracture initiates at the center of the neck and propagates outward resulting in a curved fracture surface as shown in Fig.~\ref{fig::mech1_local}(c-\textit{iii}). It is important to note that the resolution of these features, and the simulation of neck formation, growth, and fracture demand a remeshing-based continuum FE simulation. Notably, the global compressive strain is still small at $\strain = 0.05$, whereas locally the material has attained its fracture strain $\peeq_{f}$ of 2.0 under tension. 

Fig.~\ref{fig::mech1_local}(d) shows enlarged views of zones labelled `S' and `T' in Figs.~\ref{fig::mech1_local}(a-\textit{iii}) and~\ref{fig::mech1_local}(b-\textit{iii}). It depicts inter-member contact and self-contact of the folded compression members. Notably, it is seen that the regions with $\peeq \geq 1.0$  constitute only a small fraction (less than 12\%) of the entire member even under severe local deformation. The through-thickness plastic strain distribution is clearly nonlinear in the regions of high member curvature, and it is again only possible to capture this using continuum modelling.

Supplementary videos M1 and M2~\cite{supp} show the evolution of deformation, plastic strain, and the stresses in members in zones `P' and `Q' respectively, over the entire simulation. At small global strains $\strain \leq 0.03,$ plastic buckling of member `$c$' is seen to precede the tensile yield of member `$t$'; this is typical. Moreover, the plastic strain $\peeq$ in the tension member `$t$' is homogeneous only up to a value of $0.2$. 

\subsection{Truss action in macro-cell walls triggers the NB mode}
\label{subsec:load}
\begin{figure}[ht]
	\centering
	\begin{minipage}{0.77\textwidth}
		\includegraphics[width=\textwidth]{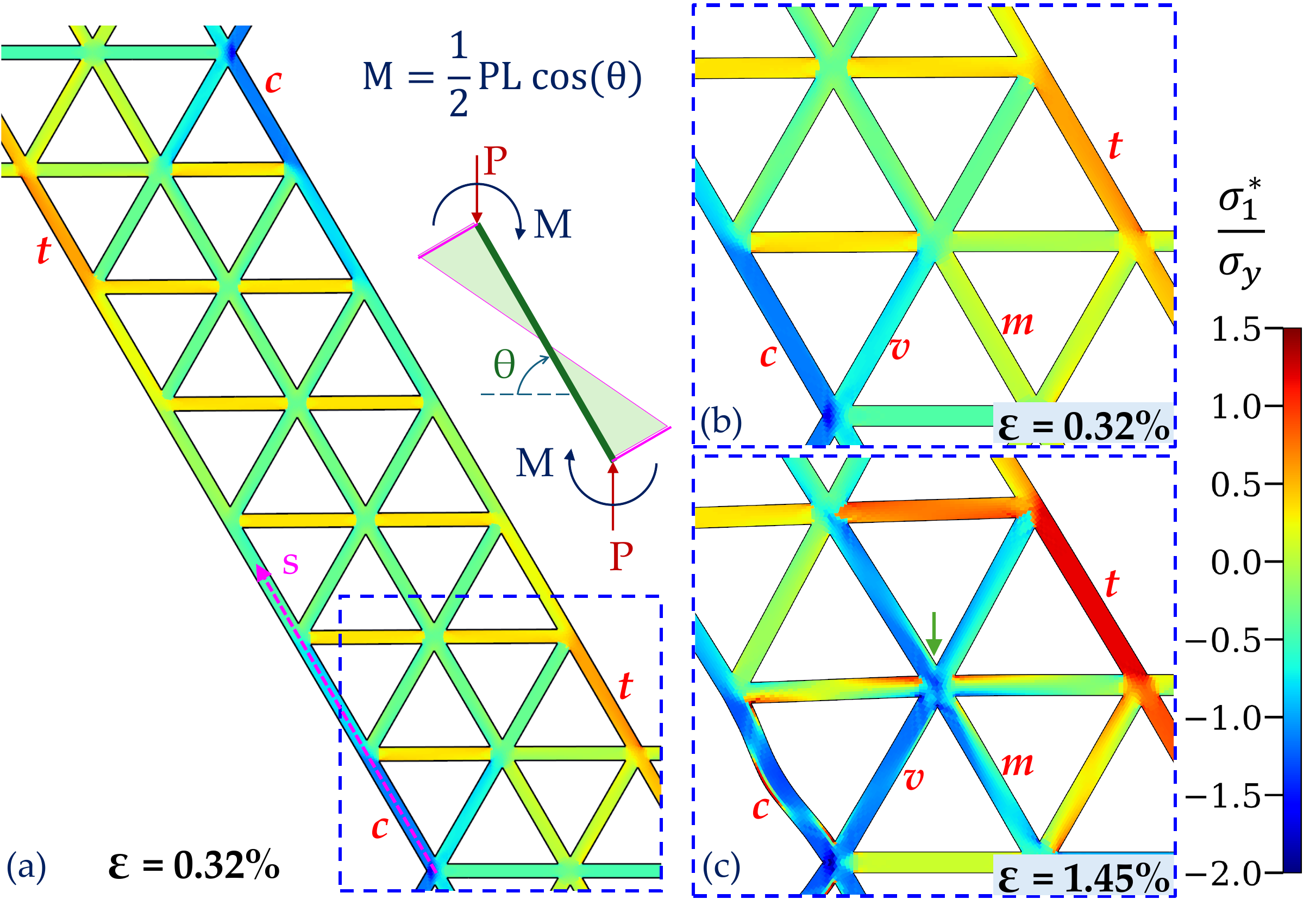}	
	\end{minipage}
	\begin{minipage}{0.22\textwidth}
		\includegraphics[width=\textwidth]{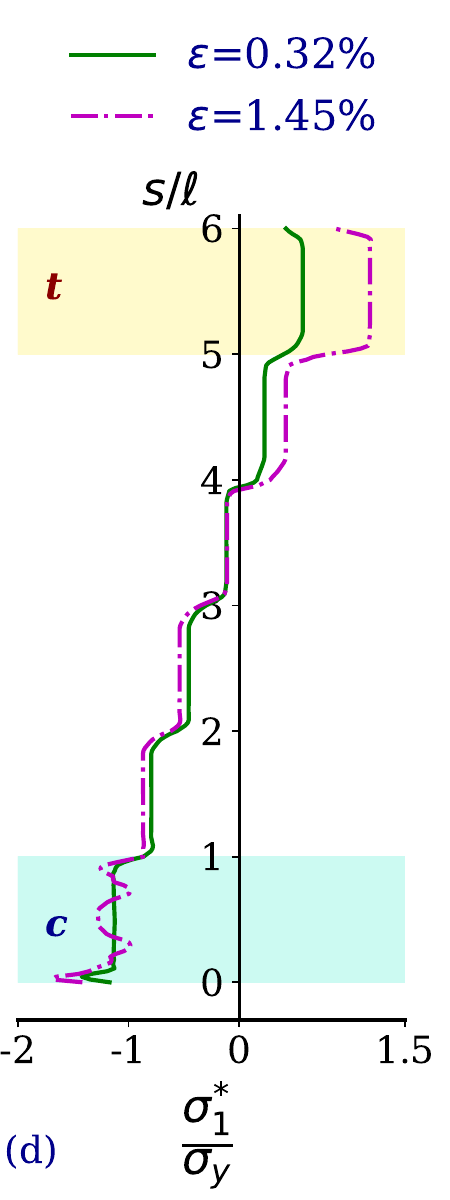}	
	\end{minipage}
	\caption{(a) Normalized stress $\sigma_{1}^{*}/\sigma_{y}$ in a representative macro-cell wall at global strain of $\strain = 0.32\%$. (b \& c) Zoomed-in views of the stress distribution ($\sigma_{1}^{*}/\sigma_{y}$) in the boxed, near-junction zone in panel (a) at global strains of $\strain = 0.32\%$ \& $\strain = 1.45\%$ respectively. (d) FE stress distribution $\sigma_{1}^{*}/\sigma_{y}$ along the s-coordinate line marked in panel (a). Here, $\rho = 0.1$, $r=7$, $\sigma_{y} = 181.5$ MPa.}
	\label{fig::loadtransfer}
\end{figure}

The occurrence of the necking-buckling (NB) mode in the HTH lattice can be explained by examining the prevailing stress state prior to its onset. Define the signed maximum absolute principal stress $(\sigma_{1}^{*})$ as 
\begin{equation}
	\sigma_{1}^{*} = \left\{ 
	\begin{matrix}
		\sigma_{1}, & \text{if } |\sigma_{1}| \geq |\sigma_{2}| \\
		\sigma_{2}, & \text{if } |\sigma_{1}| < |\sigma_{2}|
	\end{matrix}
	\right.
	\label{eq::stress}
\end{equation}
where $\sigma_{1}, \sigma_{2}$ are the in-plane principal stresses. 
Fig.~\ref{fig::loadtransfer}(a) shows the stress distribution in a typical macro-cell wall of the HTH lattice at a low global strain of 0.32\%, prior to the onset of the NB mode. The maximum absolute principal stress $\sigma_{1}^{*}$ is normalized with respect to the initial yield stress $\sigma_{y} (= 181.5$ MPa), and is plotted over the deformed geometry. Every microscale member in Fig.~\ref{fig::loadtransfer}(a) is in a nearly-uniform state of stress, suggestive of a stretching-dominated substructure. Further, the outer members (e.g. `$t$' and `$c$') on either side of the macro-cell wall constitute tension-compression pairs. The `web' region of the cell wall possesses two sets of equally-stressed members; one (diagonal, 60$^\circ$) under compression, and the other (horizontal, 0$^\circ$ in Fig.~\ref{fig::loadtransfer}(a)) under tension. This is because of truss action, wherein the outer axial  members resist the bending moment and the oblique web members carry the shear force at every section. The tension-compression pair arrangement provides an increased moment arm compared to a conventional hexagonal honeycomb (HEX) at the same relative density, and explains the reported superior performance (see~\cite{yin_2018}) of HTH over HEX. 

The inset in Fig.~\ref{fig::loadtransfer}(a), obtained by structural analysis~\cite{gibson_1997} of the macro-cell wall, predicts constant axial and shear forces accompanied by a linearly-varying bending moment that attains a maximum at the two ends. The FE stress distribution along one of the chords with respect to the $s$-coordinate in Fig~\ref{fig::loadtransfer}(a) is plotted in Fig.~\ref{fig::loadtransfer}(d); the change in the sign of $\sigma_1^*$ is consistent with the change in the sign of bending moment in elementary analysis. Due to axial compression of the cell wall, the maximum compressive stress exceeds the maximum tensile stress, and the zero stress point is shifted away from the mid-span (occurring at $s/l \approx 3.9$), nearer to  the tension end of the chord. This also explains why plastic buckling (in members like `$c$') always precedes tensile yielding (in members like `$t$'). 

Figs.~\ref{fig::loadtransfer}(b) and~\ref{fig::loadtransfer}(c) show the stress distribution in members in the boxed zone in Fig.~\ref{fig::loadtransfer}(a) before and after the plastic buckling of member `$c$', at $\strain=0.32\%$ and $\strain=1.45\%$ respectively. Prior to buckling, `$c$' and `$t$' are the maximally stressed compression and tension members respectively, and a near-constant stress state prevails in all the members in Fig.~\ref{fig::loadtransfer}(b). As seen in Fig.~\ref{fig::loadtransfer}(c), buckling of `$c$' triggers a load redistribution that results in a break-down of the uniform stress state not only in `$c$' but also in the six members connected to the junction (indicated by a green arrow). The load redistribution manifests as undulations in the corresponding stress in member `$c$' (magenta plot in Fig.~\ref{fig::loadtransfer}(d)). Load redistribution also causes   increased stress in members like `\textit{v}' and stress reversals in members like `\textit{m}'. As seen in Figs.~\ref{fig::loadtransfer}(c) and~\ref{fig::loadtransfer}(d), `$t$' continues to carry a high and uniform tensile stress despite load redistribution, causing it to neck and eventually fracture.

\subsection{Coordinated buckling (CB) of the web members deters necking}
\label{subsec:CB}
\begin{figure}[t]
	\centering
	\begin{subfigure}{0.45\textwidth}		
		\includegraphics[width = \textwidth, left]{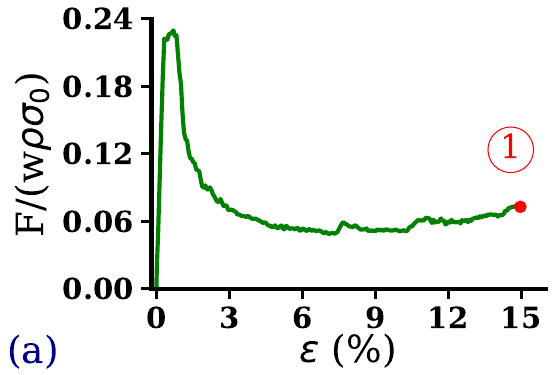}
	\end{subfigure}
	\begin{subfigure}{0.54\textwidth}
		\includegraphics[width = \textwidth, right]{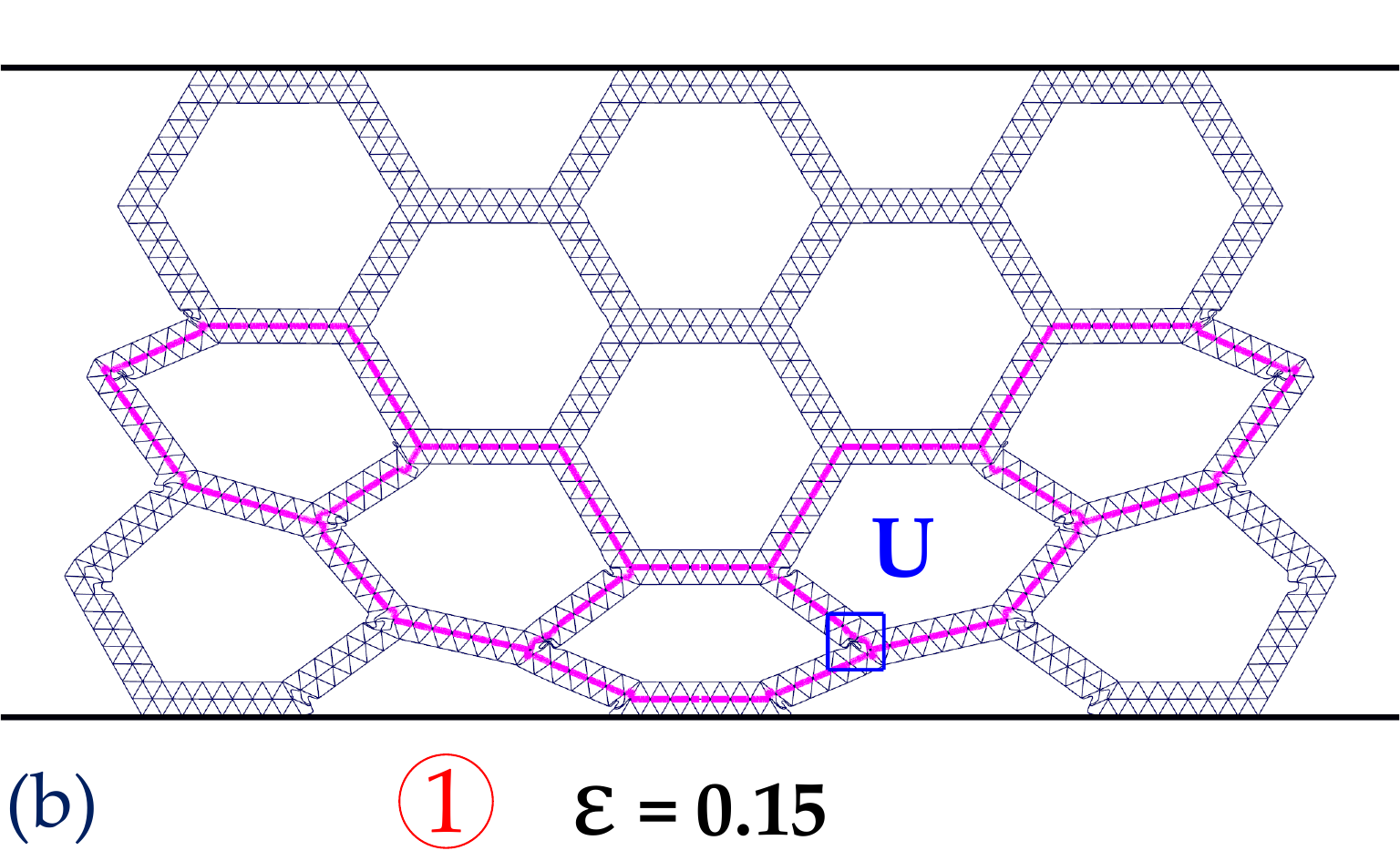}
	\end{subfigure}
	\begin{subfigure}{\textwidth}
		\includegraphics[width=\textwidth]{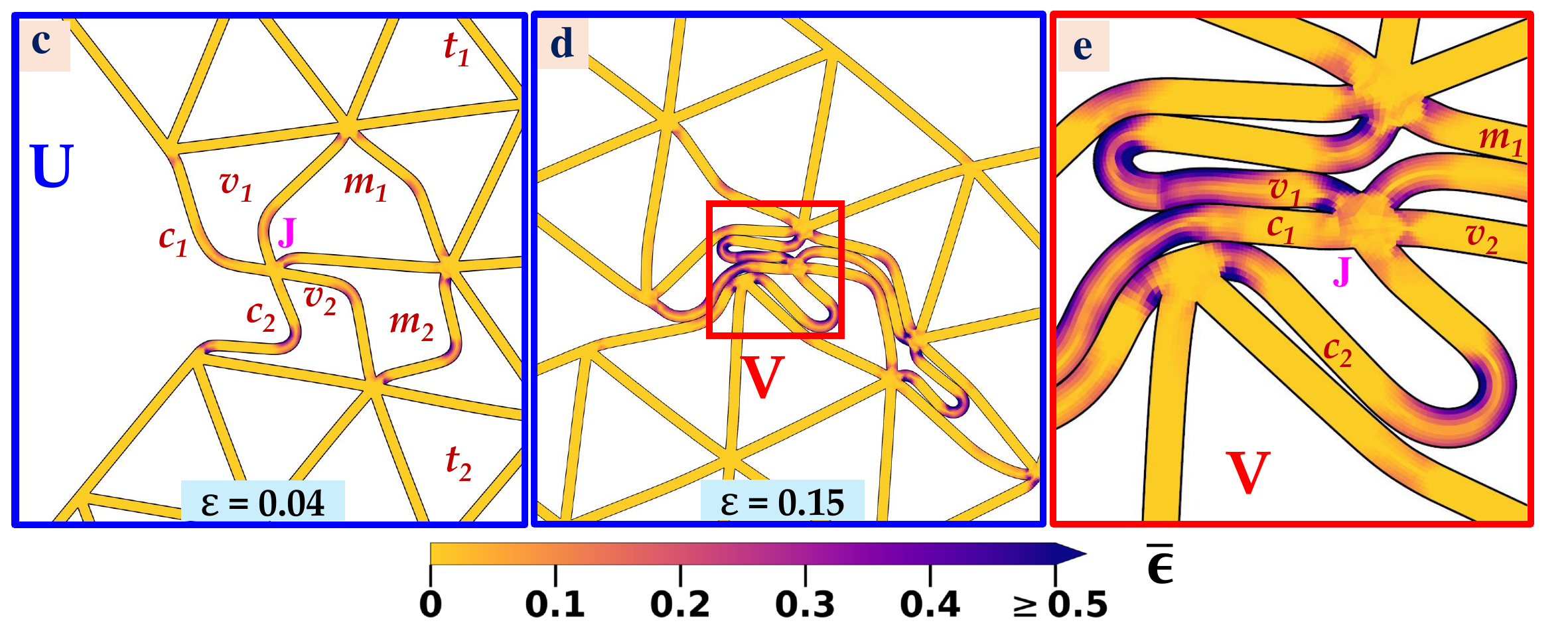}
	\end{subfigure}
	\caption{(a) Global $\force-\strain$ response of an $r=7$, $\rho=0.06$ HTH lattice subjected to uniaxial compression (b) Deformed geometry at a global strain of $\strain = 0.15$. Frame sequence (c-d) show the deformed geometry of members undergoing coordinated buckling (CB) in zone `U' in panel (b). The superimposed colour depicts $\peeq$  (e) Zoomed-in view of zone `V' in panel (d) showing inter-member contacts and self-contacts.
	The $\peeq$ range is clamped to a maximum of 0.5 for better contrast.} 
	\label{fig::mech2_global}
\end{figure}
Necking of tension members is not observed in uniaxial compression of an HTH lattice with a lower relative density of $\rho=0.06$ despite its having the same structural layout and scale ratio $r = 7$. Fig.~\ref{fig::mech2_global}(a) presents the global force-strain response of the $r=7, \rho=0.06$ HTH lattice up to a global strain of $\strain=0.15$. The initial response is linearly elastic upto a strain of $\strain = 0.3\%$, with a peak force $\force_{p} = 0.23$ attained at $\strain = 0.7\%$ as the critical compression members undergo plastic buckling. The post-critical force drop is significantly steeper than in the necking-buckling case (Fig.~\ref{fig::mech1_global}(a)).

Fig.~\ref{fig::mech2_global}(b) shows the deformed geometry corresponding to a global strain of $\strain = 0.15$. The deformation localizes in a somewhat different pattern than the $\rho=0.1$ case; only one layer of cells accommodates most of the imposed strain, as opposed to two layers in Fig.~\ref{fig::mech1_global}(b). Consequently, the deformation of a cell in the localizing zone is more pronounced. As before, actual plastic straining is further localized to the ends of macro-cell walls. 

The frames (c--d) in Fig.~\ref{fig::mech2_global} show the deformation history of members in zone `U' in Fig.~\ref{fig::mech2_global}(b) as the imposed strain $\strain$ increases from $0.04$ to $0.15$. Again, the superimposed colour depicts $\peeq$. Fig.~\ref{fig::mech2_global}(c) shows that multiple members ($c_1, v_1, m_1, c_2, v_2, m_2$) have buckled plastically, unlike the NB mode in Figs.~\ref{fig::mech1_local}(a-\textit{i})--(a-\textit{iii}) which involves buckling of the critical member `$c$' only. In fact, the central junction `J' rotates, forcing the members connected to it ($c_1, v_1, c_2, v_2$) to buckle in a coordinated fashion. The term `coordinated buckling' (CB) will thus be used to collectively refer to deformation modes that involve buckling of multiple members without any member undergoing necking. Plastic straining is confined to the ends and mid-span of these buckled members, with $\peeq$ attaining a maximum value of 0.48. Upon further loading to $\strain = 0.15$, the buckled members develop inter-member contacts and self-contacts, eventually folding upon themselves smoothly as seen in Fig.~\ref{fig::mech2_global}(d). Fig.~\ref{fig::mech2_global}(e) shows a zoomed-in view of the boxed zone `V' in Fig.~\ref{fig::mech2_global}(d). A maximum $\peeq$ of 1.08 is attained in the concave face of the folded member `$v_1$'. The extent of plastic strain localization seen in Figs.~\ref{fig::mech2_global}(d) and (e) is remarkable; even microscale members that participate in coordinated buckling (like `$v_1$' and `$c_2$') have large unyielded regions. 
Again, the importance of continuum modelling to accurately resolve the complex pattern of contacts, and the highly nonlinear, through-thickness distribution of $\peeq$ in members like `$v_1$' and `$c_1$' in Fig.~\ref{fig::mech2_global} will be realized. Supplementary video M3~\cite{supp} shows the evolution of deformation, plastic strain, and the stresses in members in zone `U' over the entire simulation for the CB mode.

\begin{figure}[h]
	\centering
	\begin{subfigure}{0.4\textwidth}
		\includegraphics[width=\textwidth]{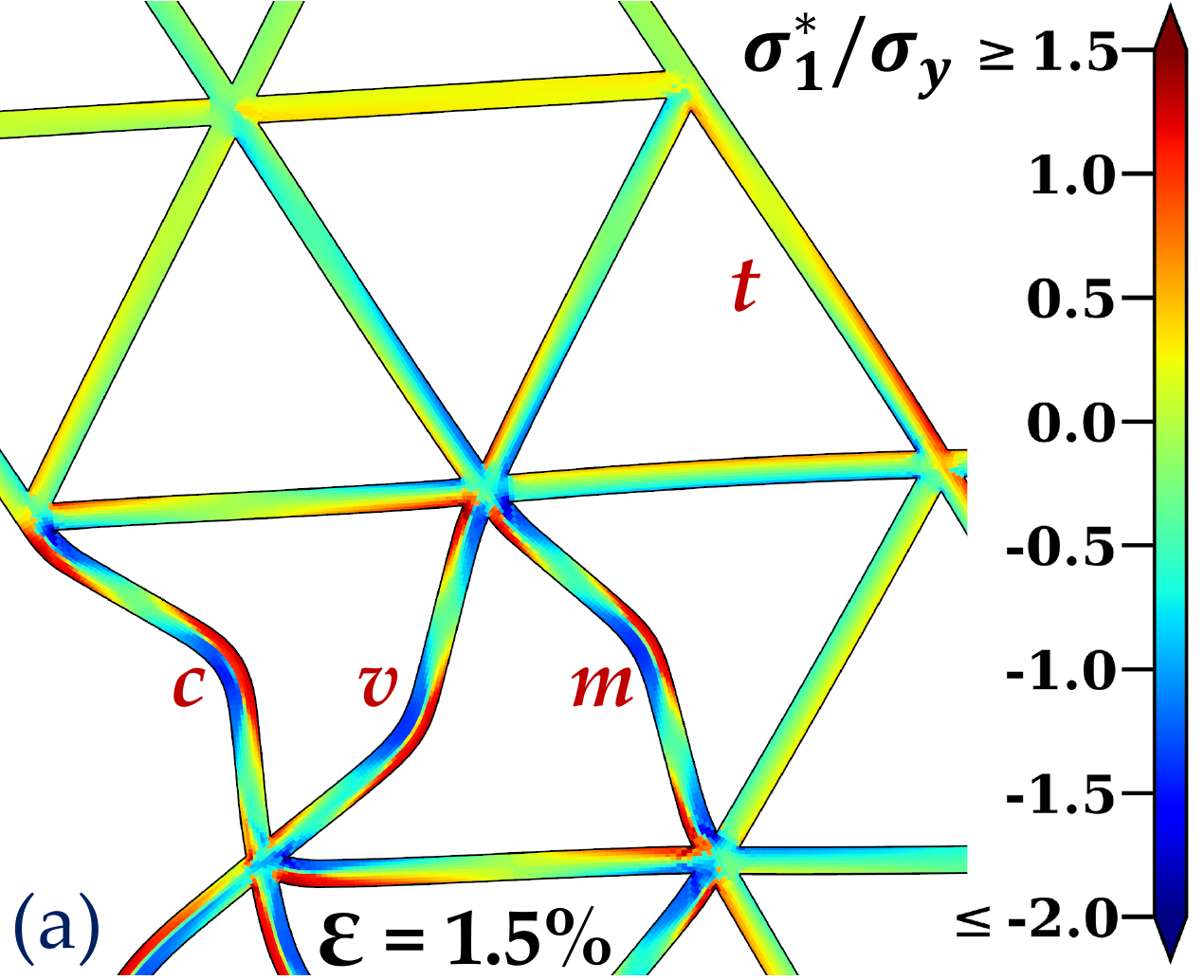}		
	\end{subfigure}
	\begin{subfigure}{0.59\textwidth}
		\includegraphics[width=\textwidth]{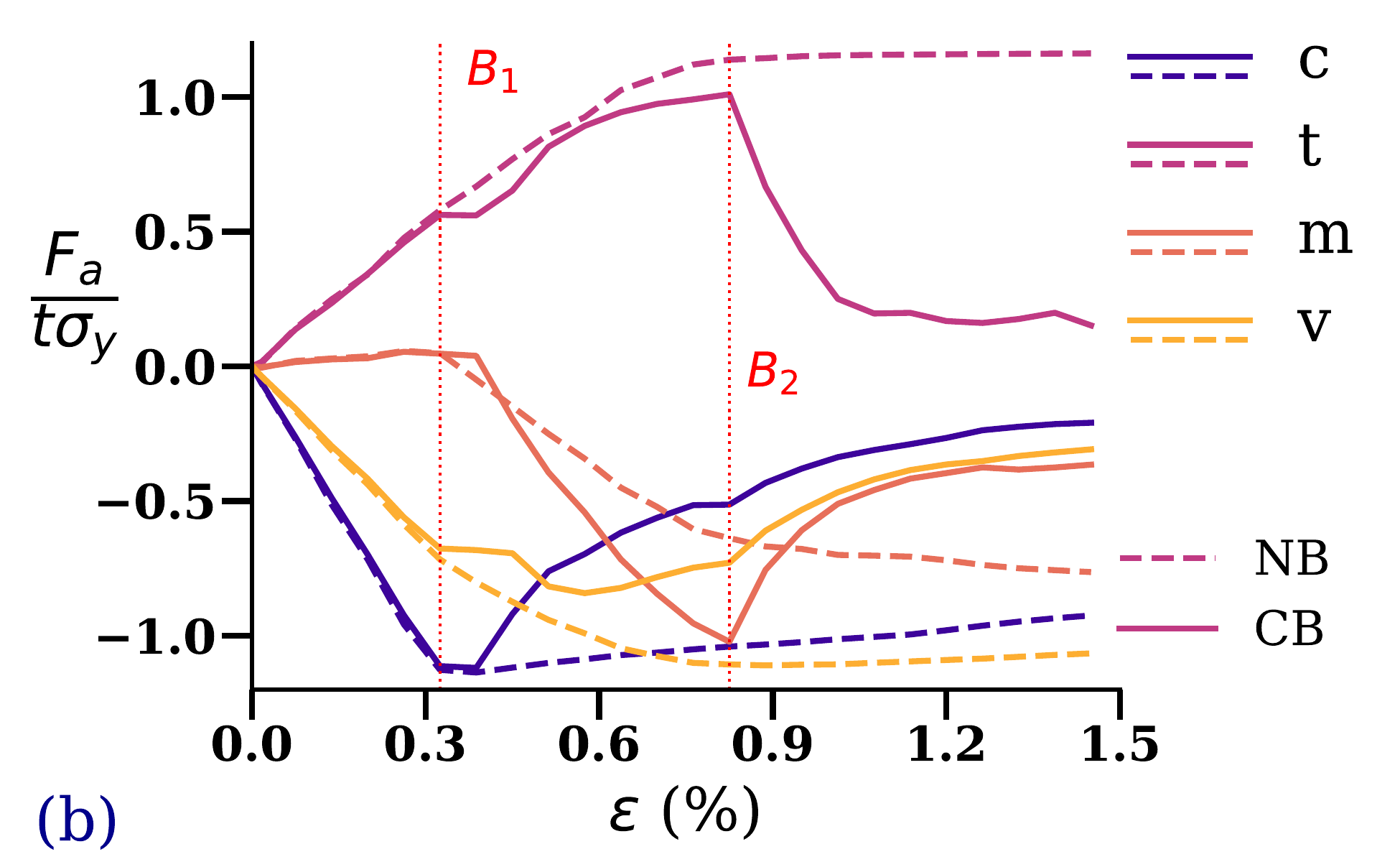}		
	\end{subfigure}
	\caption{(a)~Stress distribution $\sigma^{*}_{1}$~(see~Eq.~\eqref{eq::stress}) superimposed on the deformed geometry in a set of members participating in the CB mode. The imposed global strain is $\strain=1.5\%$ (b) Evolution of member forces ($\hat{F}_{a} = F_{a}/t\sigma_{y}$) in the CB mode (solid lines) compared with those in the NB mode (dashed lines) in Sec.~\ref{subsec:NB}. The HTH lattice showing the CB mode has $r=7, \rho=0.06$. $\sigma_y = 181.5$ MPa in both cases.}
	\label{fig::hth_rho6_stresses}
\end{figure}

\noindent
Notably, truss action prevails prior to the onset of the CB mode as well, and the CB mode is also initiated by plastic buckling of maximally stressed compression members like `$c_1$' and `$c_2$' in Fig.~\ref{fig::mech2_global}(c). However, the ensuing load redistribution occurs in way that prevents necking in the tension member `$t$'. Fig.~\ref{fig::hth_rho6_stresses}(a) shows the $\sigma_{1}^{*}/\sigma_y$ (see Eq.~\eqref{eq::stress}) distribution at a typical junction of an $r=7, \rho=0.06$ HTH lattice at a global strain of 1.5\%. %
Note that at this strain, the members `$c$', `$v$',  and `$m$' are visibly buckled, and none of them, including `$t$', is in a uniform state of stress. 

We track forces in the two sets of members marked `$c$', `$t$', `$m$', and `$v$' in Fig.~\ref{fig::hth_rho6_stresses}(a) for the CB mode and Fig.~\ref{fig::loadtransfer}(b) for the NB mode from a global strain of 0 to 1.5\%. %
Fig.~\ref{fig::hth_rho6_stresses}(b) presents the evolution of the normalized axial force $\hat{F}_{a} = F_{a}/t\sigma_{y}$ in these members with increasing global $\strain$; the corresponding member forces in a lattice that shows the NB mode (see Fig.~\ref{fig::loadtransfer}(b)) are plotted using dashed lines. Here, $F_{a}$ is the component of the sectional force resultant along the \textit{undeformed} axial direction $\mathbf{E}_{a}$ defined as 
\begin{equation}
	F_{a} = \int\limits_{S}\mathbf{E}_{a} \cdot \,{\sigb}\,\mathbf{n}\,dS
	\label{eq::force}
\end{equation}
where $\mathbf{n}$ is the normal to any section $S$, and ${\sigb}$ is the Cauchy stress tensor.

Initially, the normalized forces are identical in the NB and CB modes in each of the four members (`$c$', `$t$', `$m$', `$v$'). At a strain labelled $B_{1}$ in Fig.~\ref{fig::hth_rho6_stresses}(b), the maximally stressed member `$c$' buckles plastically in both modes at a critical $\hat{F}_{a}$ of about -1.1 (violet solid and dashed lines). Buckling of the member `$c$' triggers a load redistribution in both modes, wherein the force in member `$m$' (dark orange solid and dashed lines) switches to compressive from tensile and subsequently increases in magnitude up to a strain $B_{2}$, compensating for the force drop in member `$c$'. Note that the magnitude of the compressive force in `$m$' increases rapidly in the CB mode to attain a critical value at -1.0, which causes member `$m$' to buckle as well (see Fig.~\ref{fig::hth_rho6_stresses}(a)). This does not happen to the `$m$' member in the NB mode as its axial force remains sub-critical ($\hat{F}_{a}$ = -0.64) at the stage $B_{2}$. 

On increasing the strain from 0 to $B_2$, the normalized forces in member `$t$' increase monotonically in both modes, with the values differing only slightly at $B_2$. However, in the CB mode, there are no more non-buckled axial members to redistribute the compressive force to, at this stage. 
Therefore the tensile force in `$t$' also drops to maintain equilibrium, thereby causing a global force drop. Note that the maximum tensile force $\hat{F}_{a}$ attained in `$t$' at this stage is only about 1.0 which is inadequate to cause necking, and the force drop precludes any possibility of future neck-formation. In contrast, the NB mode does not feature a second buckling event, and the tensile force in `$t$' continues to increase above 1.2 (eventually rising high enough to cause necking at $\strain = 3\%$), while the compressive force in `$m$' remains sub-critical at 0.87. 
In summary, plastic buckling of `$m$' in the CB mode deters further increase in tensile force in `$t$' and prevents necking. 
\subsection{Local and global effects of scale ratio $(r)$ and relative density $(\rho)$}
\begin{figure}[ht]
	\centering
	\begin{subfigure}{\textwidth}
		\includegraphics[width=\textwidth]{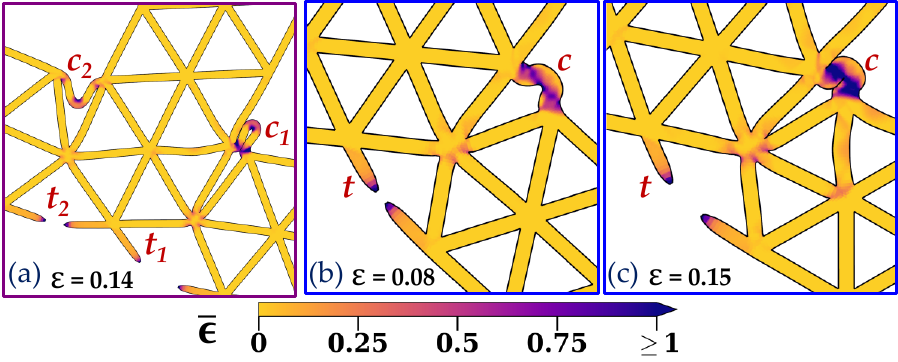}
	\end{subfigure}		
	\caption{Snapshots of the necking-buckling (NB) mode as seen in (a) A low-density $\rho = 0.06$ and $r = 11$ lattice and (b) A high-density $\rho = 0.14$ and $r = 7$ lattice. The necking and buckling members are labelled `$t$' and `$c$' respectively. Panel (a) shows two pairs of active members participating in the NB mode. Panel (c) shows cusp-formation with high $\peeq$ in the high-density lattice at a global strain of 0.15.}
	\label{fig::param_local}
\end{figure}
The necking-buckling (NB) mode is seen in HTH lattices over a range of densities and scale ratios. 
For instance, Fig.~\ref{fig::param_local}(a) shows the deformed geometry of a low-density ($\rho = 0.06, r = 11$) HTH lattice at a global $\strain$ of 0.14, with the superimposed colour depicting $\peeq$. Evidently, the NB mode remains the principal mechanism enabling rotations of the macro-cell walls about the vertices of the hexagonal macro-cells. Interestingly, as the deformation is more localized than the $\rho = 0.1$ case, multiple NB pairs ($(c_1, t_1)$ and $(c_2, t_2)$ in Fig.~\ref{fig::param_local}(a)) are formed in the vicinity of the junctions, thus accommodating greater relative rotation between macroscale beams. The buckled member `$c_1$' folds in upon itself forming a self-contact. 

A high-density ($\rho = 0.14, r=7$) lattice also shows the NB mode; one such junction with NB is shown in Fig.~\ref{fig::param_local}(b) at $\strain = 0.08$. However, due to the low slenderness of the microscale members ($\lambda$ = 6.7 compared to $\lambda = 16$ in the low-density case), the member `$c$' does not smoothly fold in upon itself at high strains; instead it forms a cusp and establishes a complete self-contact as shown in Figs.~\ref{fig::param_local}(c). Again, a remeshing-based FE simulation is critical to capture this mode of deformation.

\begin{figure}[h]
	\centering
	\includegraphics[height=0.2\textheight]{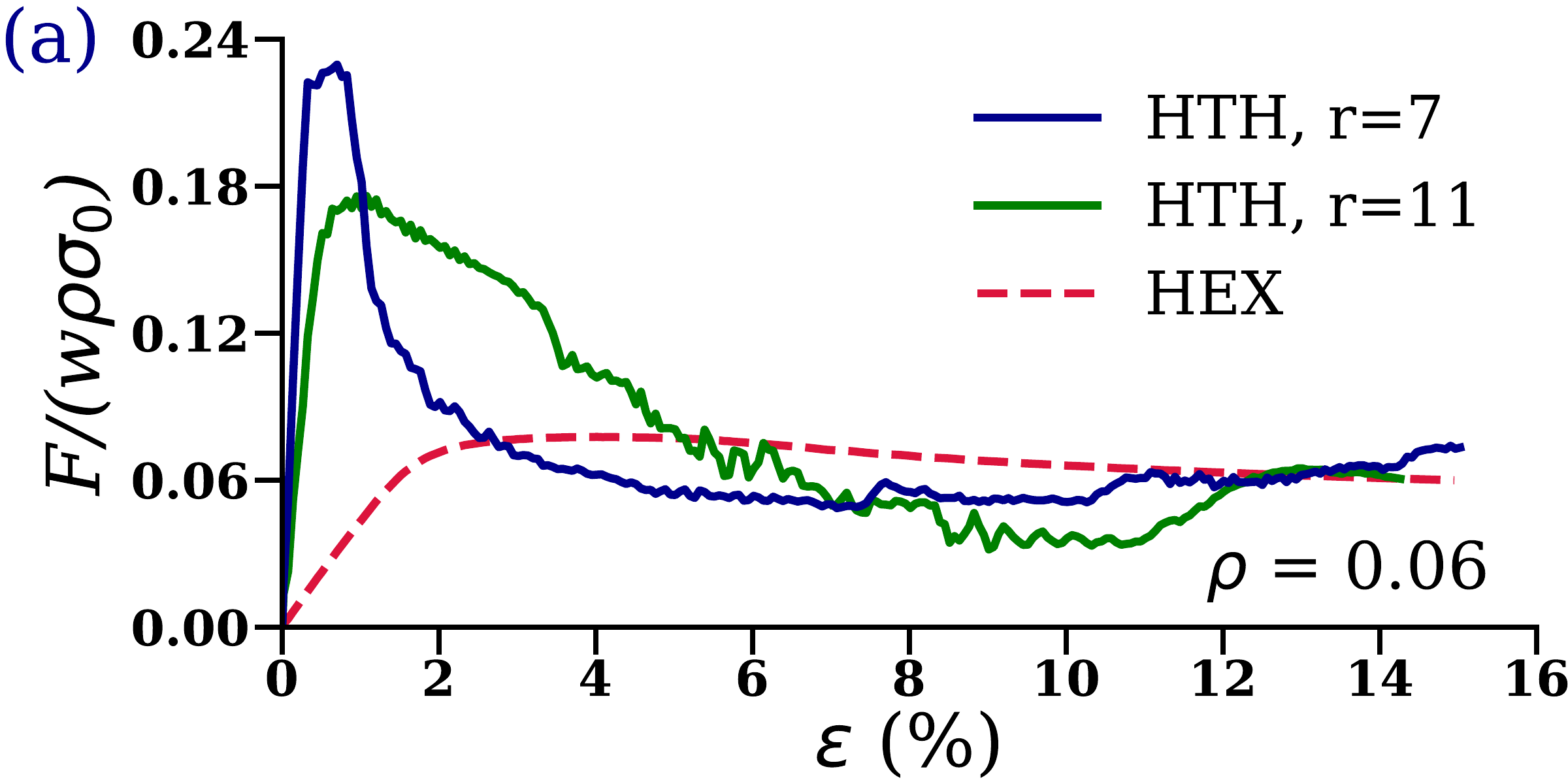}
	\includegraphics[height=0.2\textheight]{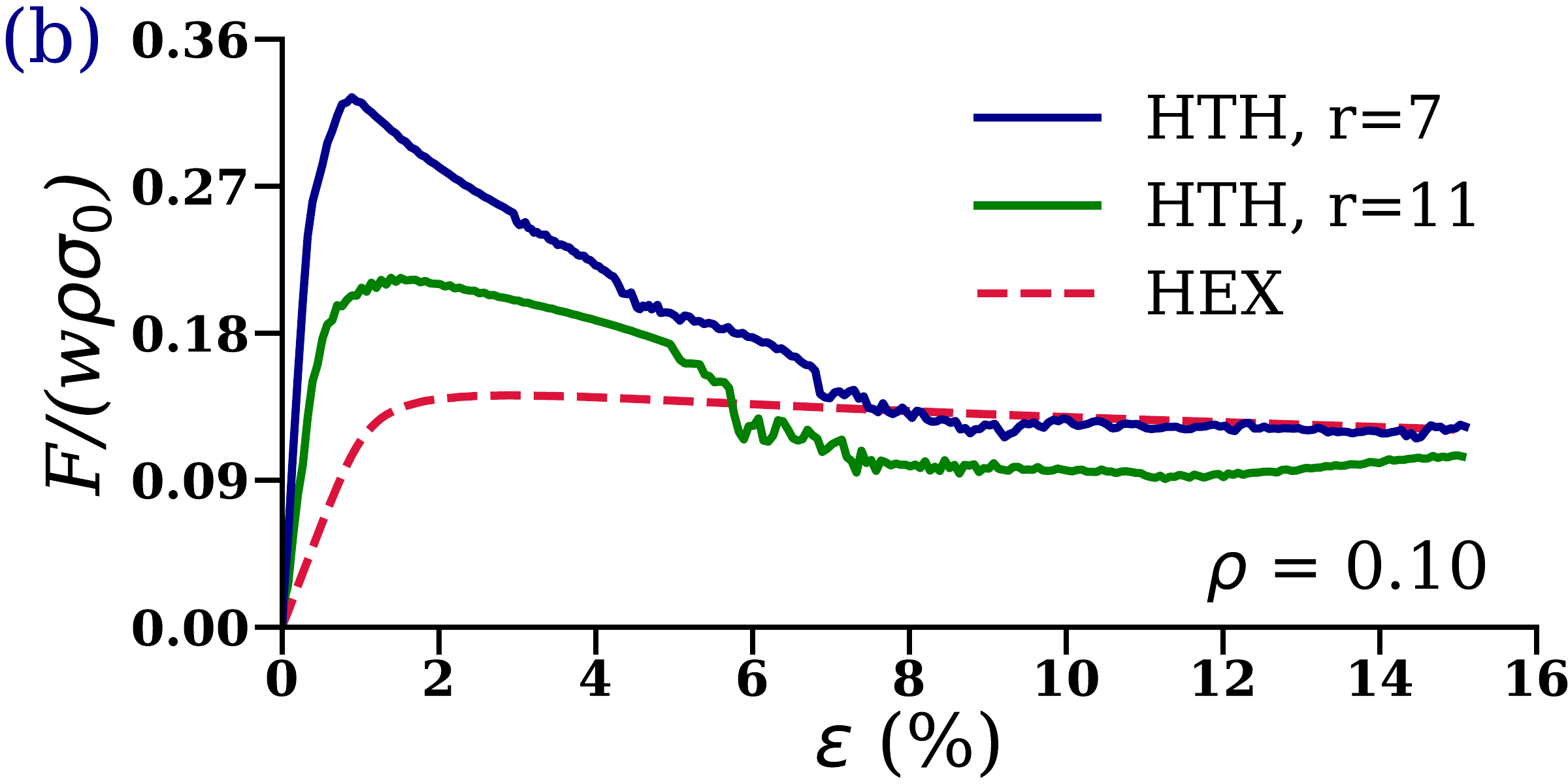}	
	\includegraphics[height=0.2\textheight]{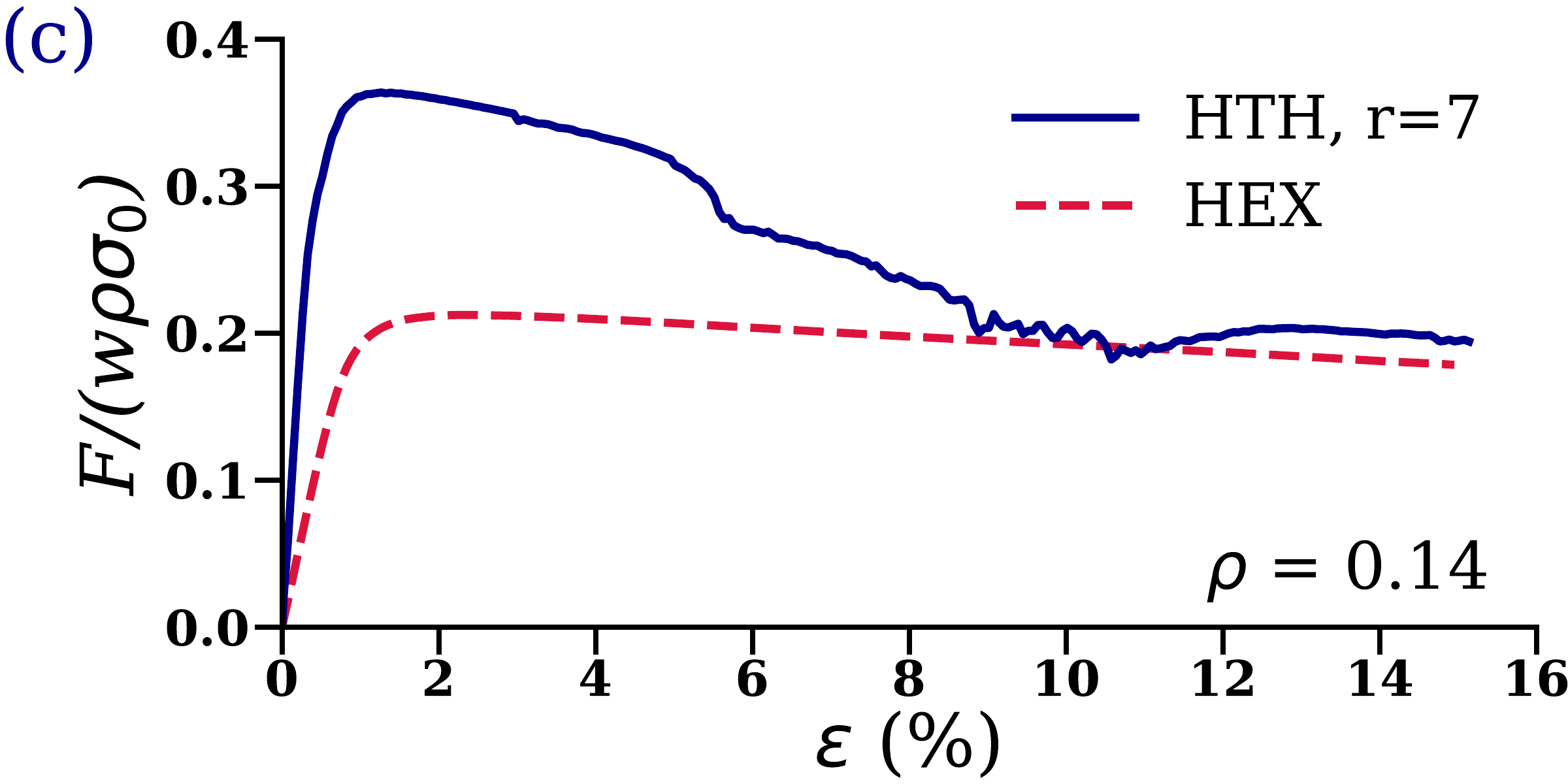}
	\caption{Global force--strain ($\force-\strain$) response of HTH lattices with (a) $\rho=0.06$  (b) $\rho=0.10$ (c) $\rho=0.14$ and the indicated $r$ values. The $\force-\strain$ responses for equivalent single-scale lattices (labelled `HEX') are also plotted.} 
	\label{fig::param_global}
\end{figure}

The uniaxial compression responses for HTH lattices with relative densities $\rho$ of 0.06, 0.10, and 0.14 are presented in Figs.~\ref{fig::param_global}(a-c). The dotted line denotes the force response of a conventional (HEX) honeycomb at the same $\rho$. 
\noindent
As seen in Fig.~\ref{fig::param_global}(a), the peak force is highest for an $r=7$ lattice, reduces for an $r=11$ lattice, and is the lowest for the conventional HEX honeycomb (with all three having $\rho = 0.06$). While both HTH lattices show significant post-peak softening due to plastic buckling, there is only marginal softening in the HEX honeycomb. In fact, all HTH lattices, regardless of density (see Fig.~\ref{fig::param_global}(b) and Fig.~\ref{fig::param_global}(c)) show significant post-peak softening and then attain a plateau at global strains between 6\% and 12\%.

\noindent
Some of the small--moderate amplitude force fluctuations seen in the response of HTH lattices are due to the release of kinetic energy that accompanies fracture of necked regions when the NB mode is active. 
For instance, in Fig.~\ref{fig::param_global}(a), the green curve, which is the $\force-\strain$ response for an $r=11$ lattice with an active NB mode, is visibly `noisier' than the $r=7$ (blue) curve, which is CB-only.
Lastly, the areas under the $\force-\strain$ curves for all the HTH lattices in Fig.~\ref{fig::param_global} are visibly higher than the area under the $\force-\strain$ curves for the corresponding HEX lattices; this indicates that HTH lattices have generally superior energy absorption characteristics, as discussed in Sec.~\ref{subsec:energy}.
\subsection{Other hierarchical lattices exhibit the necking buckling mode}
\begin{figure}
	\centering
	\begin{subfigure}{0.49\textwidth}
		\includegraphics[width=\textwidth]{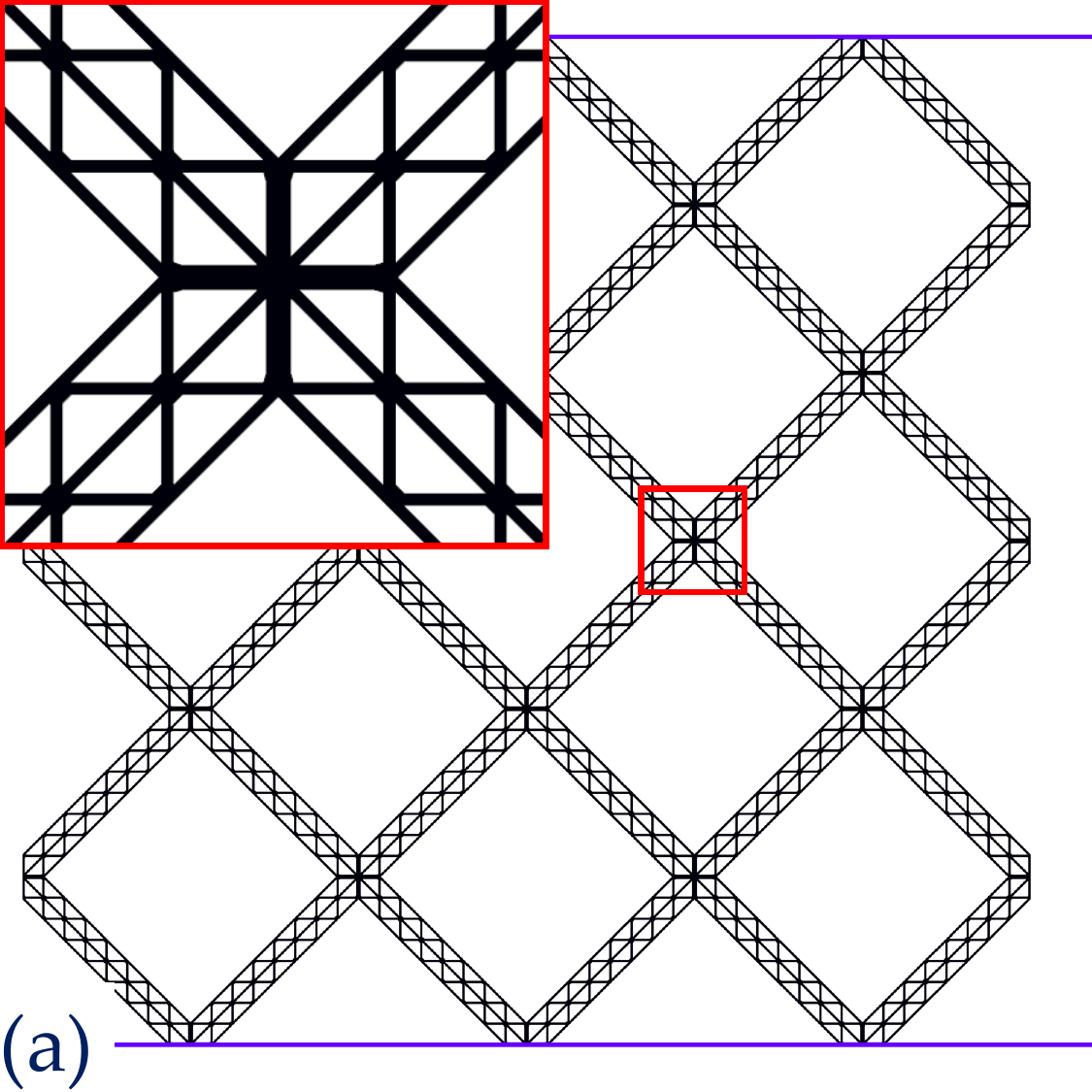}
	\end{subfigure}	
	\begin{subfigure}{0.49\textwidth}
		\includegraphics[width=\textwidth]{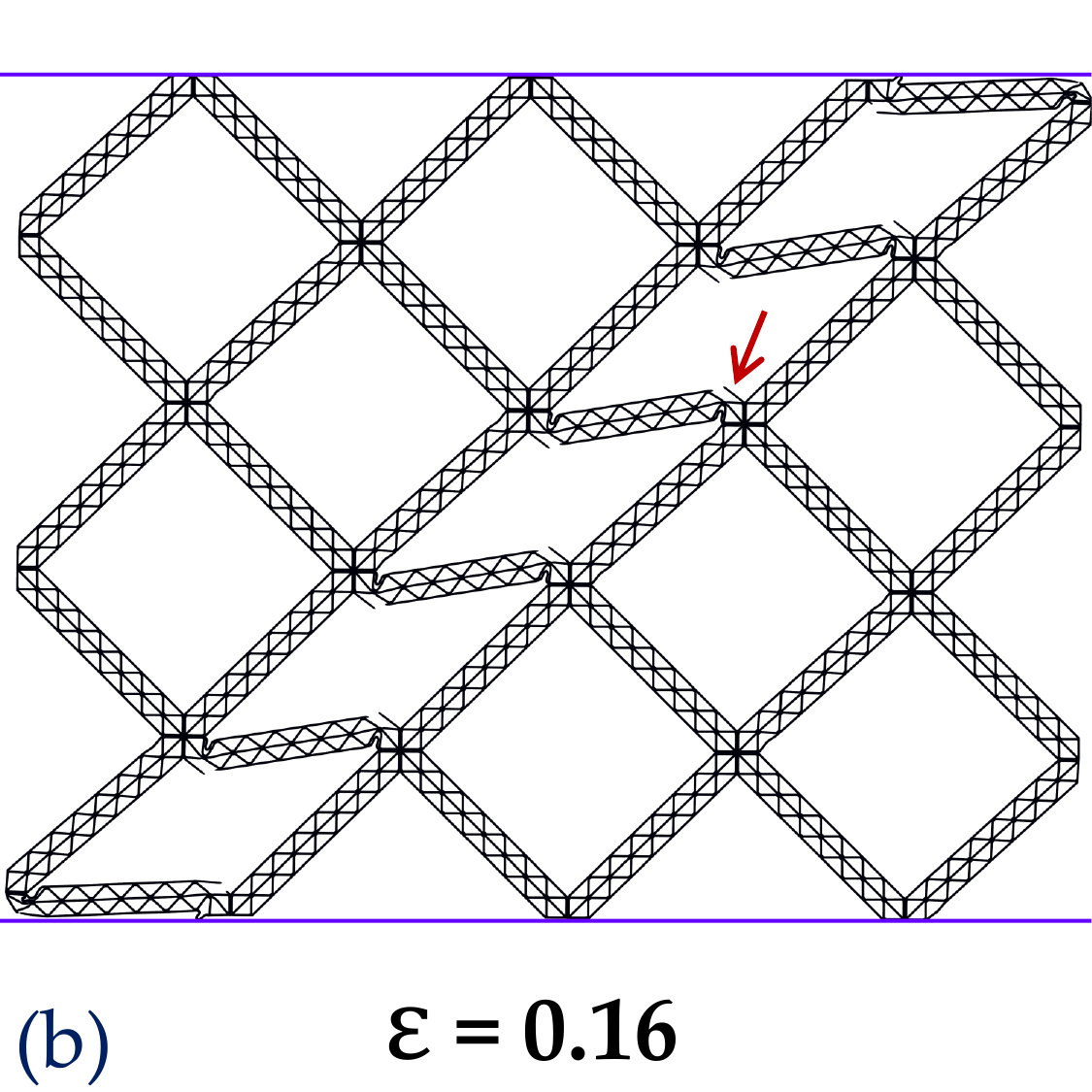}
	\end{subfigure}
	\vspace{0.3 cm}	
	\\
	\begin{subfigure}{0.49\textwidth}
		\includegraphics[width=\textwidth]{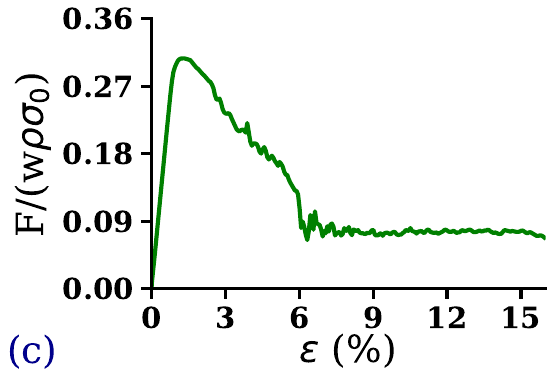}
	\end{subfigure}
	\begin{subfigure}{0.49\textwidth}
		\includegraphics[width=\textwidth]{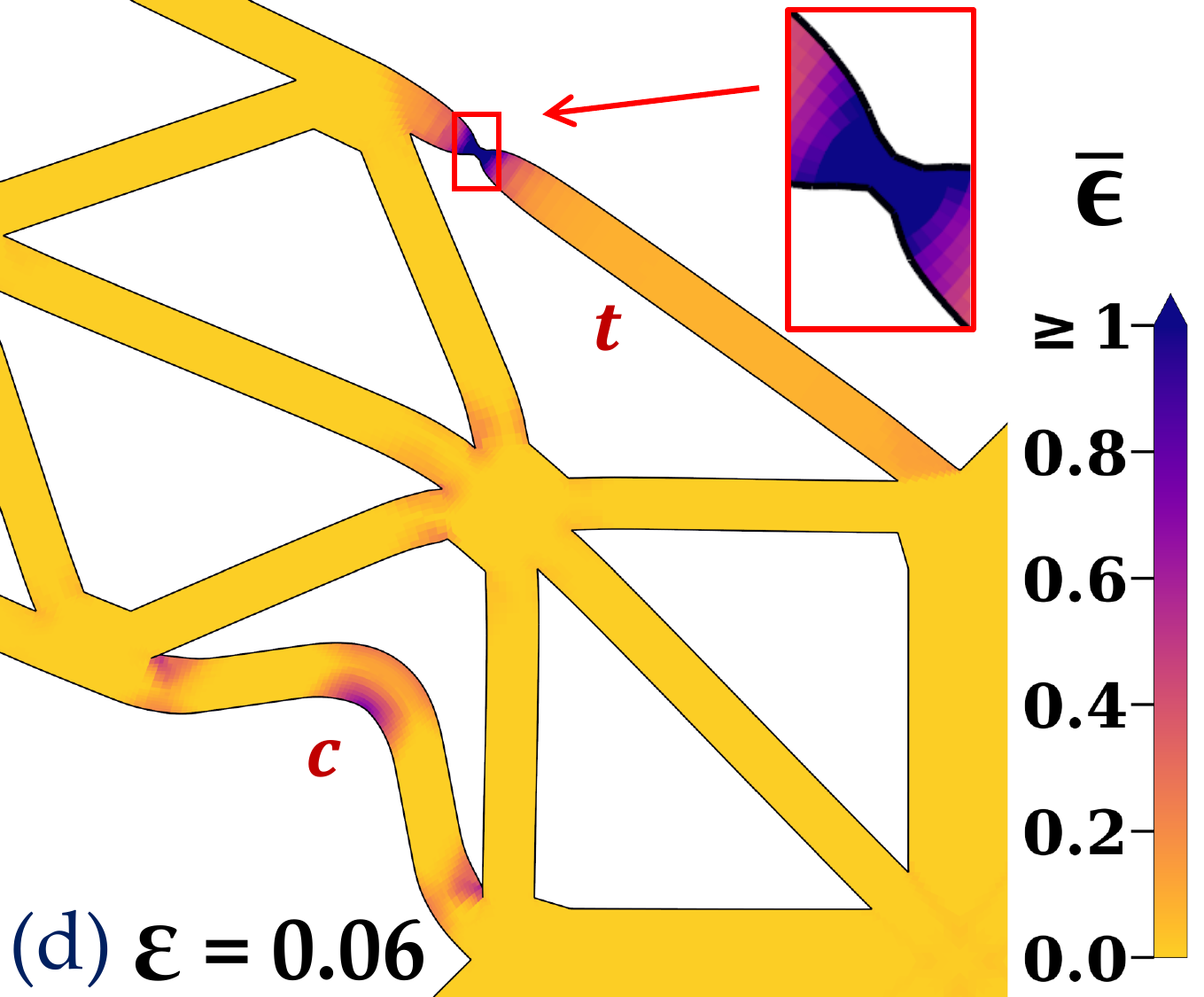}
	\end{subfigure}	
	\vspace{0.3 cm}	
	\caption{(a) Geometry of the diamond-triangle hierarchical (DTH) lattice. The inset shows a zoomed-in view of a macroscale junction (b) Deformed geometry of an $r=8, \rho = 0.10$ DTH lattice at $\strain = 0.16$ (c) Global $\force-\strain$ response of this DTH lattice under uniaxial compression (d) One frame ($\strain = 0.06$) showing the occurrence of the NB mode near the junction indicated by the arrow in (b). The participating members are labelled `$c$' and `$t$'. Note that necking occurs away from the mid-span of `$t$'. } 
	\label{fig::dth}
\end{figure}
It is important to note that the occurrence of the NB mode is not restricted to the HTH lattice, since it is a generic mode of accommodating relative rotation of truss-like macro-beams in two-scale hierarchical solids. To demonstrate this, we consider a diamond-triangle hierarchical (DTH) lattice. While geometrically very different (see Fig.~\ref{fig::dth}(a)) from the HTH lattice, it also has a bending-dominated macro-structure and a triangular substructure. A $3\times 3$ specimen of DTH with $\rho = 0.1$ and $r=8$ is subjected to uniaxial compression as shown in Figs.~\ref{fig::dth}(a) and~\ref{fig::dth}(b) . The vertical and horizontal members at the macroscale junction (inset) are thickened to prevent buckling inside the junction.

Like the HTH lattices, the global $\force-\strain$ response for the DTH lattice exhibits an initial peak followed by steady softening to a plateau as shown in Fig.~\ref{fig::dth}(c). Fig.~\ref{fig::dth}(d) shows a zoomed-in view of a typical junction (indicated by the arrow in Fig.~\ref{fig::dth}(b)). The NB mode is clearly visible, with the members labelled `$c$' and `$t$' undergoing buckling and necking respectively. 
Unlike HTH lattices, the location of the neck is not near the mid-span, but closer to one end of the tension member. This is possibly due to the fact that the other end is connected to a thicker member. 

\section{A theoretical model to predict the NB and CB buckling modes}
\label{sec:theory}
The remeshing-based, fully-nonlinear FE analyses in Section~\ref{sec:results} accurately simulate the elastic-plastic response of two-scale hierarchical solids, revealing the NB and CB buckling modes.  However, it is also desirable to have a simplified theoretical model to quickly elucidate the underlying drivers for these modes. In other contexts, the two-scale homogenization model~\cite{qiao_2016} and its variants have been successful in estimating the collapse load of a variety of hierarchical lattices~\cite{qiao_2016, li_2020, wang_2021, liu_2022, xu_2022, zhang_2022}. 

\begin{figure}
	\centering
	\includegraphics[width=\textwidth]{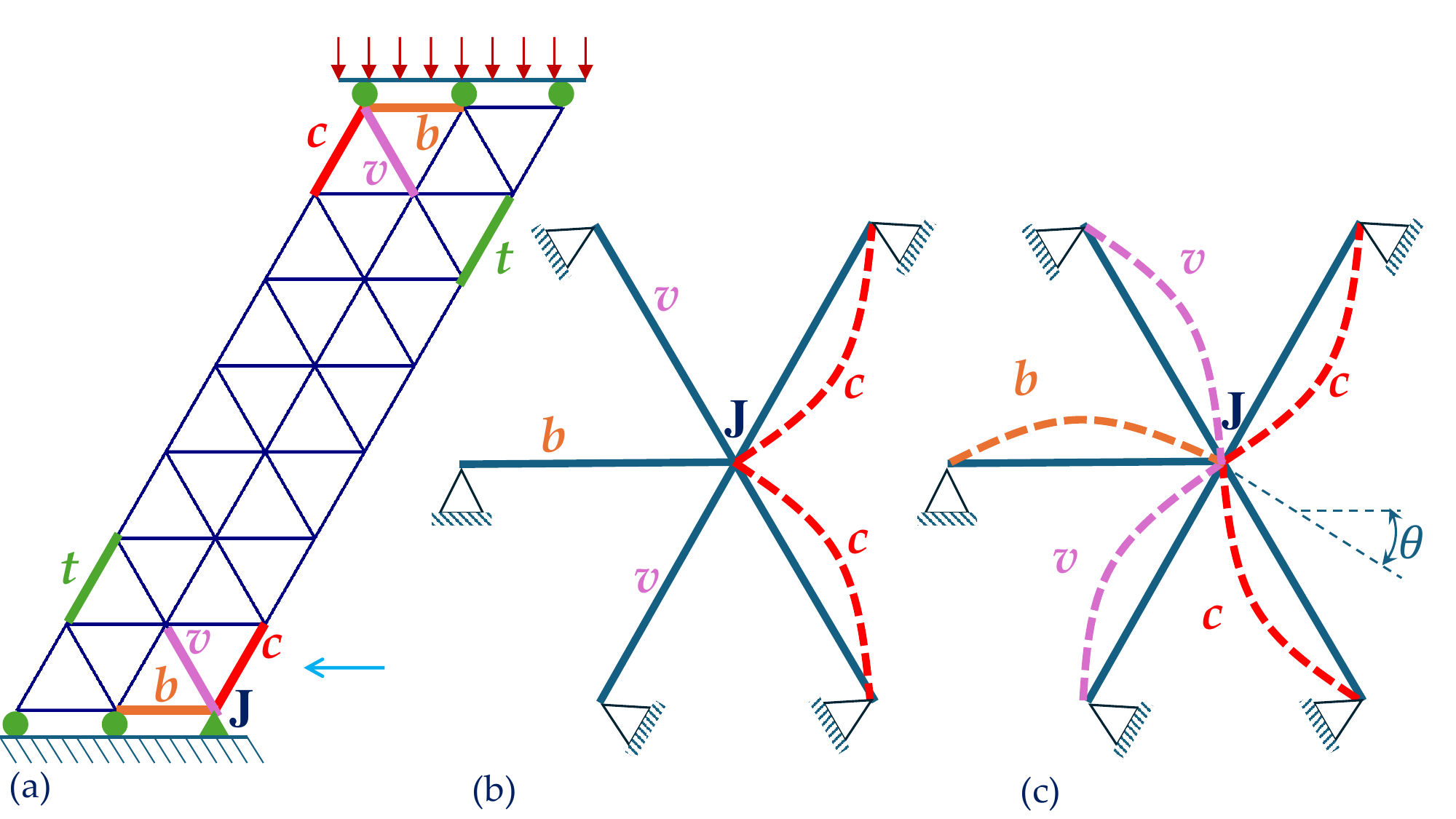}		
	\caption{(a) Schematic of a typical HTH macro-cell wall used for the theoretical analysis in Sec.~\ref{sec:theory}, with the critical members highlighted. Panels (b) \& (c) show the mode shapes of the members participating in the NB and CB modes respectively, at junction `J'.}
	\label{fig::theory}
\end{figure}

\subsection{Linearised elastic analysis of a macro-cell wall}
As discussed in Sec.~\ref{subsec:load}, truss action prevails in the macro-cell wall because the microscale members are axially loaded prior to the onset of the NB or CB modes. At this stage the deformation is very small, and one can therefore use the undeformed configuration for the analysis. 
The macro-cell wall is thus treated as a truss fixed at the bottom with a uniform compressive displacement imposed on it, as shown in Fig.~\ref{fig::theory}(a) for an example truss. 
Given the high degree of static indeterminacy, the Integrated Force Method~\cite{patnaik_1973} is employed to obtain the microscale member forces $\{\mathbf{F}\}$. For a truss with $n$ degrees of freedom and $m$ members, the assembly of force equilibrium equations for all non-zero degrees of freedom (d.o.f) is expressed in matrix-vector form as follows:
\begin{equation}
	\underset{n\times m}{\vB}\,
	\underset{m\times 1}{ \vF}
	 = \underset{n\times 1}{\vP}
	\label{eq::equil}
\end{equation}
Here $\vB$ is the (rank-deficient) coefficient matrix, $\{\mathbf{P}\}$ is the vector of applied nodal forces, and $s = m - n$ is the degree of static indeterminacy. The general solution to Eq.~\eqref{eq::equil} is given as~\cite{patnaik_1973, przemieniecki_1985} 
\begin{equation}
	\underset{m\times 1}{\vF}
	 = \underset{m\times 1}{\vFs{0}} + 
	 \underset{m\times s}{\vBs{1}}\,
	  \underset{s\times 1}{\vR}
	\label{eq::gensol}
\end{equation}
where $\vFs{0}$ is a particular solution that satisfies $\vB \vFs{0} = \vP$, $\vR$ is the vector of redundant forces, and the columns of $\vBs{1}$ constitute a basis for the nullspace of $\vB$. Physically, the $i^{th}$ column of $\vBs{1}$ gives the member forces in the primary structure under the application of a unit force in the $i^{th}$ redundant. Therefore, the incompatibility $\Delta_{R}^{i}$ of the $i^{th}$ redundant member can be calculated using the unit virtual load method~\cite{przemieniecki_1985} as follows:
\begin{equation}
	\delta W_{ext}  = \delta W_{int} \implies 1\, \cdot \, \Delta_{R}^{i}  =\vBss{1}{i}^{T}\big(\vG\,\vF\big)
	\label{eq::virt}
\end{equation}
Here $\vG$ is the $m\times m$ diagonal matrix of member flexibilities. The compatibility conditions (for $i=1, 2, ..., s$) are therefore expressed in matrix-vector form as follows: 
\begin{equation}
	\delR = \vBs{1}^{T}\,\vG\,\vF = \{\bom{0}\}
	\label{eq::comp}
\end{equation}

Next, we enforce the non-zero nodal displacement boundary conditions using the unit load method~\cite{przemieniecki_1985}; one can relate the vector of specified nodal displacements $\delP$ to the member forces $\vF$ as follows:
\begin{equation}
	\underset{3\times 1}{\delP}
	=
	\underset{3\times m}{\left[\dfrac{\partial \bom{F}}{\partial \bom{P}}\right]^{T}}
	\,
	\underset{m\times m}{\vG}\,
	\underset{m\times 1}{\vF}
	\label{eq::dispbc}
\end{equation}
Here the $j^{th}$ column of $ \left[\frac{\partial \bom{F}}{\partial \bom{P}}\right]$ gives the member forces due to a unit virtual load $P^{j} = 1$. Thus, the complete set of equations for the member forces is obtained from Eqs.~\eqref{eq::equil},~\eqref{eq::comp}, and~\eqref{eq::dispbc} as:
\begin{equation}
	\begin{split}
		& 
		\underset{m\times m}{
		\begin{bmatrix}
			\vB \\
			\vBs{1}^{T}\,\vG 
		\end{bmatrix}
		}
		\,
		\underset{m\times 1}{
		\vF
		}
		 = 
		\underset{m\times 1}
		{
		\begin{bmatrix}
			\vP\\
			\{\bom{0}\}
		\end{bmatrix}
		}
	\end{split}
\end{equation}
subject to the constraint Eq.~\eqref{eq::dispbc}.
Using the elimination procedure in~\cite{patnaik_1973}, the matrices $\vBs{1}$ and $[\frac{\partial \bom{F}}{\partial \bom{P}}]$ are now computed. 

\noindent
The results of elastic analyses of HTH macro-cell walls for different $r$ and $\rho$ using the above procedure show that the member marked `$c$' in Fig.~\ref{fig::theory}(a) is maximally stressed at all scale ratios $r\geq 2$. This implies that any failure mode of HTH necessarily initiates in member `$c$', as is also the case with the NB and CB modes.

\subsection{Elastic buckling analysis}
\begin{figure}[h!]
	\centering
	\includegraphics[width=0.75\textwidth]{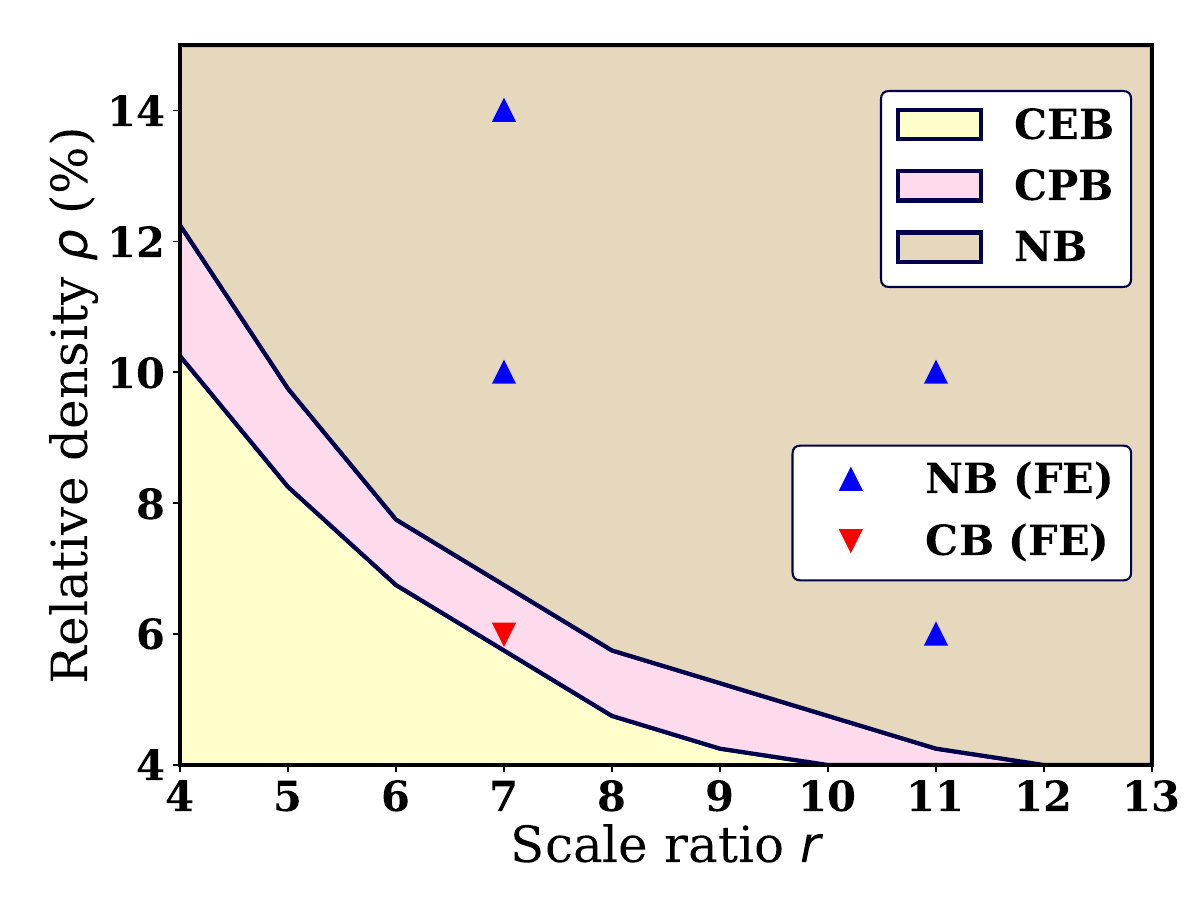}%
	\caption{Theoretically obtained buckling mode map for HTH lattices with $4\leq r \leq 13$ and $0.04 \leq \rho \leq 0.15$. Here, the necking-buckling mode is denoted by NB, and the coordinated buckling mode is denoted by CEB and CPB when the participating members buckle elastically and plastically respectively. The superimposed triangular markers depict the parameters chosen for the FE simulations.}
	\label{fig::theory_results}
\end{figure}

Schematics of the NB and CB buckling modes are shown in Figs.~\ref{fig::theory}(b) and~\ref{fig::theory}(c) respectively. As noted previously in Sec.~\ref{subsec:CB}, the key difference between the modes is that the central junction `J' rotates in the CB mode, forcing the connected members (`$c$', `$v$', `$b$') to buckle in a coordinated fashion; whereas, in the NB mode, only the `$c$' members buckle, with the junction `J' and other connected members remaining intact. By modelling the microscale members as beam columns, one can conduct a linearised buckling analysis, with the lateral deflection $w$ governed by~\cite{bazant_2010}:
\begin{equation}
	EI\,w''''(x) + P\,w''(x) = 0
	\label{eq::beamcolumn}
\end{equation}
where $P>0$ is the compressive force in the member and $EI$ is its flexural rigidity. All far-end junctions are idealized as hinged supports to obtain conservative estimates of the buckling load. Elastic buckling requires the rotations to be compatible at `J', and hence only the CB mode is feasible\footnote{It is known that modes that maximize the effective length of buckling members are preferred, see~\cite{fan_2009}}. Consequently, the boundary conditions for elastic buckling of a microscale member of length $\ell$ are 
\begin{equation}
w(0) = w(\ell) = w''(\ell) = 0 \quad,\quad w'(0) = \theta 
\label{eq:boundaryconditions}
\end{equation}
where the junction `J' has $x=0$. The solution to this boundary value problem is given by:
\begin{equation}
	w(x) = \dfrac{\theta (\ell-x)\sin(k\ell)}{k\ell\cos(k\ell) - \sin(k\ell)} + \dfrac{\theta\ell\sin(k(\ell-x))}{k\ell\cos(k\ell) - \sin(k\ell)}
\end{equation}
where $k^{2} = P/EI$.
Since elastic loading is proportional till the critical point, a load factor $\gamma$ is introduced. The force in the member `$c$' is set as $\gamma\, t\sigma_{y}$, and the other member forces are proportionately adjusted. 
The bending moment $M_{a}$ in a generic member `$a$' at the junction `J' ($x=0$) is given by:
\begin{equation}
	M_{a}(\gamma) = EI_{a}\,w_{a}''(0) = \dfrac{EI_{a}\, \,\theta k_{a}^2 \ell_{a} \sin(k_{a}\ell_{a})}{k_{a}\ell_{a}\cos(k_{a}\ell_{a}) - \sin(k_{a}\ell_{a})}; \hspace{0.5 cm} k_{a} = \sqrt{P_{a}(\gamma)/EI_{a}}
	\label{eq::moment}
\end{equation}
The load factor $\gamma_{e}$ for elastic buckling is obtained by solving the following characteristic equation which results from the equilibrium of moments at junction `J'.
\begin{equation}
	2\,M_{c}(\gamma_{e}) +  2\,M_{v}(\gamma_{e}) + M_{b}(\gamma_{e}) = 0
\end{equation}
If $\gamma_{e} < 1$, the set of members undergo coordinated elastic buckling (CEB); instead, if $\gamma_{e} \geq 1$, the member `$c$' yields prior to buckling. The set of parameters $(r, \rho)$ of HTH lattices with CEB as the critical failure mode is shaded in yellow in the buckling-mode map in Fig.~\ref{fig::theory_results}. It is the preferred mode at low density and low scale ratios.

\subsection{Post-yield stability analysis}
The post-yield stability analysis is simplified by assuming a deflected shape for the plastically buckled beam-column along the lines of Je\v{z}ek~\cite{bleich_1952}, because the sectional force-moment-curvature relationship becomes nonlinear. The buckled member `$c$' in Fig.~\ref{fig::theory}(c) is assumed to take the profile $\tilde{w}(x)$ given by:
\begin{equation}
	\tilde{w}(x) = \dfrac{\theta \ell}{\pi}\sin\left(\dfrac{\pi x}{\ell}\right)
\end{equation}
which satisfies the boundary conditions in \eqref{eq:boundaryconditions}, but is not an equilibrium solution. Suppose that the assumed deflection $\tilde{w}$ is caused by a perturbing moment $\Delta M$ at junction `J'; we now examine the stability of the configuration $\theta = 0$ when the member `$c$' has yielded. Further assuming that the member forces in `$c$', `$v$', and `$b$' remain constant when a small perturbation $\theta$ is introduced, static equilibrium requires:
\begin{equation}
	\Delta M = M_{st} - M_{dst}
\end{equation}
where the stabilizing moment $M_{st} = 2M_{v} + M_{b}$ is the elastic resistance from the members $v$ and $b$, and the destabilizing moment $M_{dst} = 2P_{c}w_{max}$. Here it is assumed that the flexural resistance of the yielded column `$c$' is negligible, and the elastic moments $M_{v}$ and $M_{b}$ are calculated from Eq.~\eqref{eq::moment}. Further, one has $P_c = t\sigma_{y}$ and $w_{max} = \theta \ell/\pi$. Finally, on performing the analysis and obtaining the value of $\Delta M$, the conditions of stability can be stated as follows:
\begin{itemize}
	\item If $\Delta M < 0$, then the $\theta = 0$ configuration is unstable, and the CB mode is favoured. The parametric region in the $(r, \rho)$ space for which coordinated plastic buckling (CPB) is the likely mechanism is shaded light pink in Fig.~\ref{fig::theory_results}.
	\item If $\Delta M > 0$, then the $\theta = 0$ configuration is stable, and the NB mode is likely. These parametric regions are shaded brown in Fig.~\ref{fig::theory_results}.
\end{itemize}
Fig.~\ref{fig::theory_results} shows that the NB mode is favoured at relatively large values of scale ratio ($r$) and density ($\rho$). The superimposed markers are the FE simulations; evidently, the mode predicted by the present theoretical analysis agrees with those obtained using FE. 
Even in the CPB mode, the member `$v$' in Fig.~\ref{fig::theory} buckles elastically (not plastically) as a result of post-yield destabilization. This is substantiated by the temporal evolution of axial force in the member `$v$' in Fig.~\ref{fig::hth_rho6_stresses}(b) -- the magnitude of the maximum force in `$v$' is less than both its yield load and its Euler buckling load. Note that while the buckling mode map (BMM) in Fig.~\ref{fig::theory_results} is for $E = 69$ GPa and $\sigma_{y} = 181.5$ MPa, a similar analysis can be carried out to obtain BMMs for other metallic materials.
\section{Discussion}
The simulation results and the theoretical model in Sec.~\ref{sec:theory} indicate that the combined necking-buckling (NB) mode is an intrinsic mode of deformation in ductile, two-scale hierarchical solids (THCS). Further evidence in support of this view is provided in the current section. Moreover, the occurrence of the NB mode has crashworthiness implications, as explored below.
\subsection{Persistence of the NB mode with reduced tensile ductility}
The effective plastic strain at failure $\epsilon_f^T$ under tension is a material parameter which is alloy and heat-treatment dependent. A value of $\epsilon_f^T = 2.0$ has been used in the simulations thus far, which is representative of an annealed, ductile metal. However, it is important to note that the NB mode persists even with considerably lower $\epsilon_f^T$. 

\begin{figure}[h!]
	\centering
	\begin{subfigure}{0.5\textwidth}
		\includegraphics[width=\textwidth]{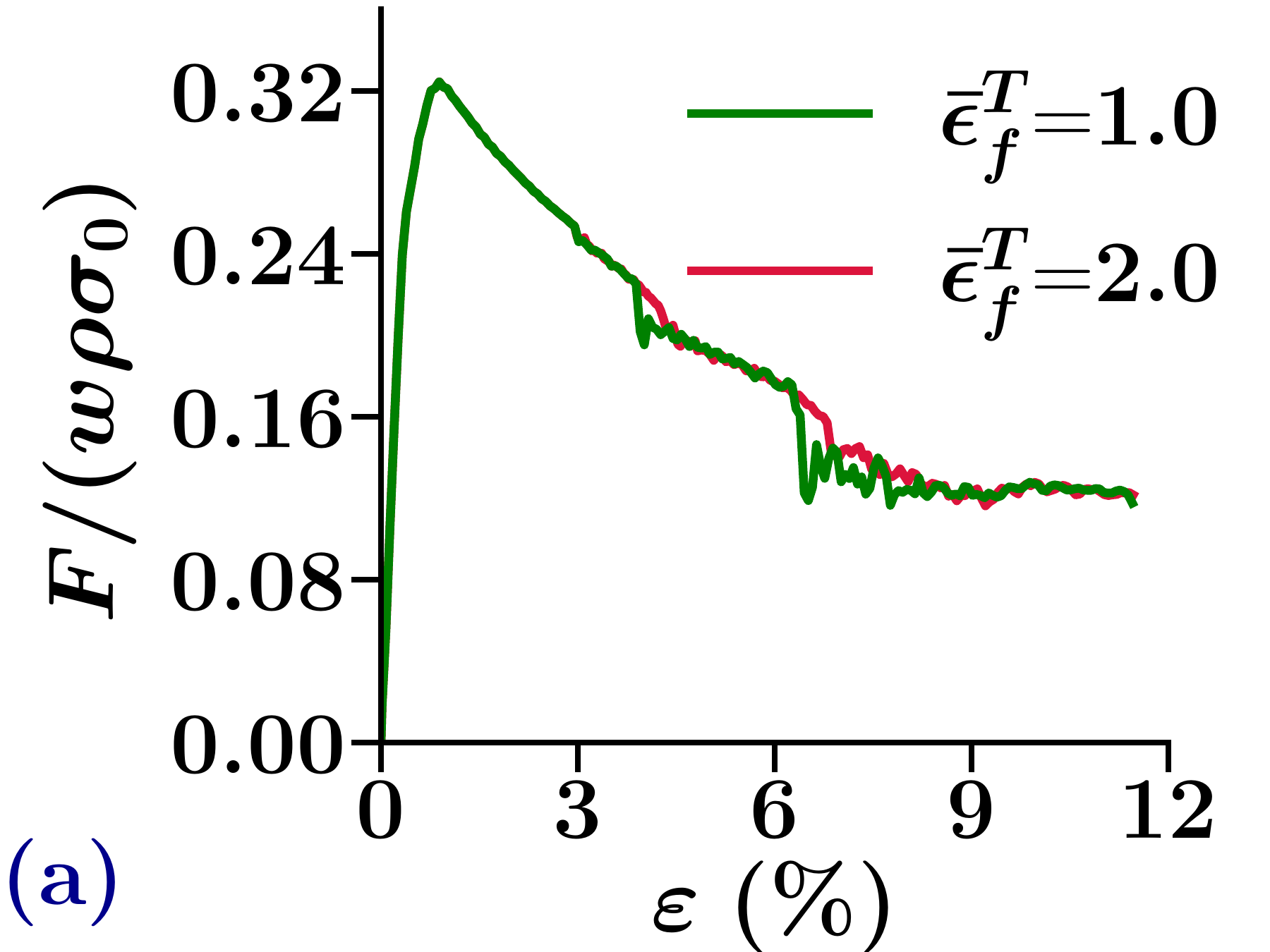}
		\vspace{6 pt}
	\end{subfigure}
	
	\begin{subfigure}{0.8\textwidth}
		\includegraphics[width=\textwidth]{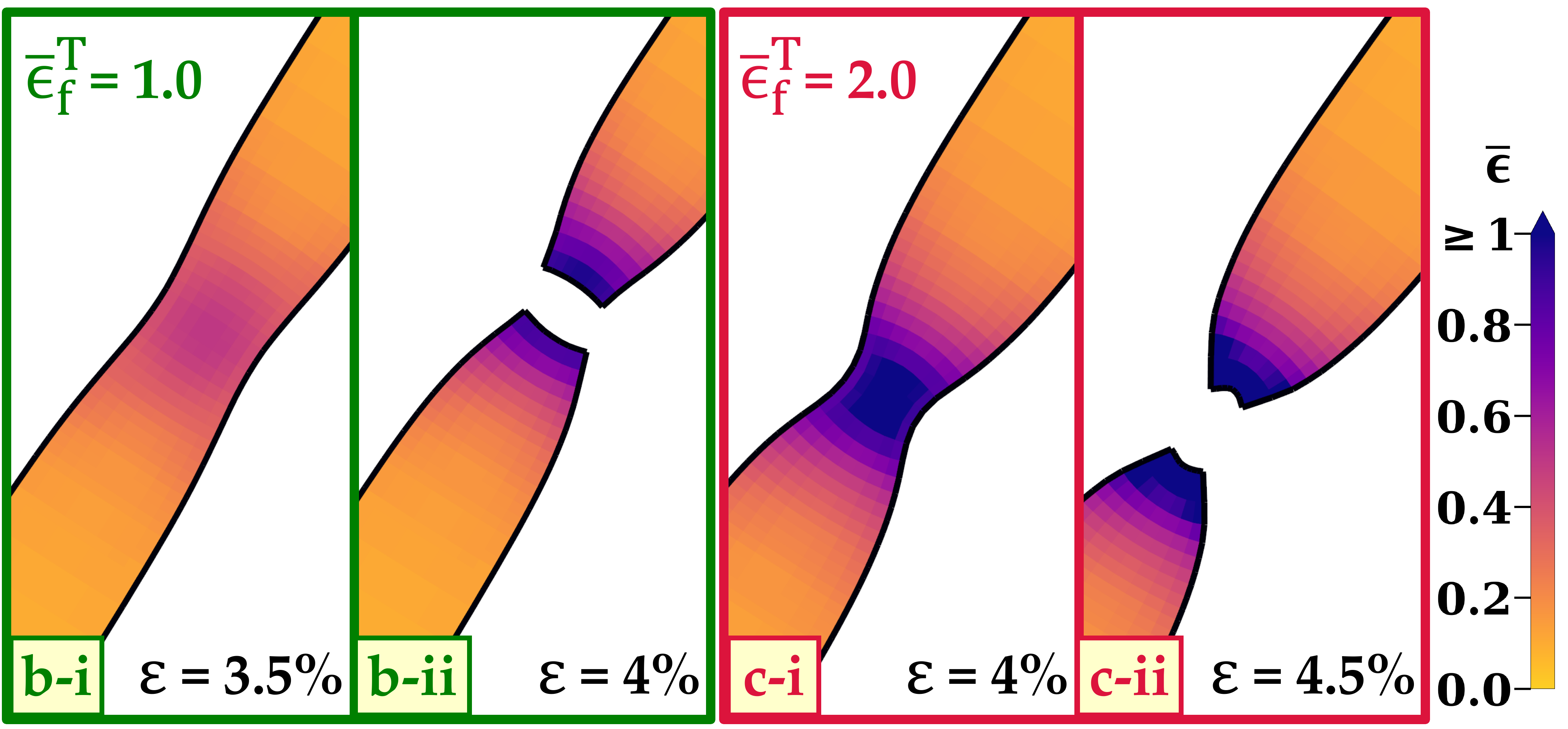}
	\end{subfigure}	
	\caption{(a) Comparison of the $\force-\strain$ responses of the $\rho = 0.1, r = 7$ HTH lattice with tensile fracture strains of $\peeq_f^{T} = 2.0$ and $\peeq_{f}^{T}=1.0$. Note the sudden force drops in the lattice with a reduced $\peeq_f^T$ of 1.0. 
	(b) With lower ductility ($\peeq_f^T = 1.0$), a tension member (i) develops less curvature and (ii) fractures earlier.  (c) With higher ductility ($\peeq_f^T = 2.0$), one sees (i) significant neck curvature and (ii) final fracture that occurs at somewhat higher global strains.} 
	\label{fig::material}
\end{figure}

Fig.~\ref{fig::material} shows the results of a uniaxial compression simulation of an HTH lattice with $\rho = 0.1, r = 7$; this lattice is identical in all respects to the HTH lattice in Sec. \ref{subsec:NB}, except that it has a reduced $\peeq_f^T$ of 1.0. Fig.~\ref{fig::material}(a) compares the $\force-\strain$ curves for the two lattices (with the green line representing the lower ductility material). It is evident that the global responses of the two lattices are very similar, except at the points corresponding to the fracture of the neck, where a sudden load drop and dynamic force oscillation is observed in the less ductile material. 

However, as long as failure in tension is preceded by necking, the NB mode itself is insensitive to the value of $\epsilon_f^T$. Fig.~\ref{fig::material}(b-\textit{ii}) shows that the fracture of the neck occurs at a lower global strain of $\strain = 0.04$ for the low ductility metal; when the ductility is higher, the final fracture occurs at $\strain = 0.045$ as seen in Fig.~\ref{fig::material}(c-\textit{ii}). Moreover, the extent of curvature developed at the neck is visibly lower for the low ductility alloy (Fig.~\ref{fig::material}(b-\textit{i})) than the high-ductility alloy (Fig.~\ref{fig::material}(c-\textit{i})).  

\noindent
If $\epsilon_f^T$ is reduced further to 0.6, necking and fracture in member `$t$' still follows the plastic buckling of `$c$'. However, at a very low $\epsilon_f^T$ of 0.3, the tension side of the buckling member `$c$' begins to undergo damage before the onset of necking in `$t$'. 
\subsection{Implications of the occurrence of the NB mode for energy absorption}
\label{subsec:energy}
The suitability of hierarchical cellular solids for energy absorption applications has been assessed using crashworthiness metrics like the peak crushing force (PCF), mean crushing force (MCF), and specific energy absorption (SEA)~\cite{li_2020, liu_2022, zhang_2022}. 
Here we use PCF, MCF, and specific plastic dissipation (SPD) (denoted by $\force_{p}$, $\force_{m}$, and $W_{p}$ respectively) to quantify the performance of various HTH lattices. They are defined as follows:
\begin{align}
	\force_{p} &= \max\limits_{0 \leq \strain \leq \strain_{f}} \force(\strain) \\
	\force_{m} &= \dfrac{1}{\strain_{f}}\int\limits_{0}^{\strain_{f}} \force(\strain)\,d\strain \\
	W_{p} &= \dfrac{1}{m}\int\limits_{0}^{t_{f}} \int\limits_V \sigb:\bom{\dot{\epsilon}}^{p}\,dV\,dt = \dfrac{1}{m}\int\limits_{0}^{t_{f}}\int\limits_{V} \bar{\sigma} \, \dot{\peeq}\,dV\,dt
\end{align}
Here $\strain_{f}$ is the final global strain at time $t = t_{f}$, $m$ is the mass of the solid, $ \bom{\dot{\epsilon}}^{p}$ is the plastic strain rate tensor, $\bar{\sigma}$ is the von Mises stress, and $\dot{\peeq}$ is the effective plastic strain rate. We use SPD rather than SEA as used by~\cite{chen_2018, zhang_2022} in order to discount the (recoverable) elastic strain energy. The PCF, MCF, and SPD are computed up to a strain of $\strain_{f} = 0.15$ for HTH lattices (over a range of $\rho$ and $r$) and for equivalent single-scale honeycombs (HEX) at the same density. Table~\ref{tab::results} summarizes these quantities for various HTH lattices, normalized by the corresponding values for HEX lattices. 

\begin{table}[h!]
	\centering
	\caption{Deformation modes, normalized energy absorption and crashworthiness parameters of HTH lattices.}
	\label{tab::results}
		\begin{tabular}{|c| c c c |c |c c c|}
			\midrule[1pt]
			Case & {$\rho$} & {$r$} & {$\lambda$} & {Mode} & {PCF*}  &  {MCF*} & {SPD*} \\
			\midrule[1pt]
			1 & 0.06 & 7 & 16.9 & CB & \textbf{2.95} & 1.11 & 1.22\\
			\hline
			2 & 0.06 & 11 & 10.8 & NB & 2.26 & 1.23 & 1.26\\
			\hline
			3 & 0.10 & 7 & 9.8 & NB & 2.28 & \textbf{1.35} & \textbf{1.54}\\	
			\hline	
			4 & 0.10 & 11 & 6.1 & NB & 1.51 & 1.04 & 1.13\\
			\hline
			5 & 0.14 & 7 & 6.7 & NB & 1.71 & 1.34 & 1.50\\
			\midrule[1pt]
		\end{tabular}
	\\
	\footnotesize{* Ratio of HTH and HEX values at the same $\rho$. For example, PCF = ${\force_{p}^{\text{HTH}}}/{\force_{p}^{\text{HEX}}}$}
\end{table}

\noindent
A low PCF coupled with high MCF and SPD are considered desirable~\cite{zhang_2022}. It is seen in Table~\ref{tab::results} that all HTH lattices exhibit increased MCF and SPD as compared to the corresponding HEX lattice. In particular, HTH lattices at $\rho=0.10, r=7$ and $\rho=0.14, r = 7$ offer the greatest enhancements in SPD($\approx 54\%$) and MCF ($\approx 35\%$) over a HEX lattice. However, these improvements are accompanied by somewhat higher PCF.
An HTH lattice with $\rho = 0.10$ and higher scale ratio of $r=11$ offers only marginal enhancements in MCF and SPD, although it has the lowest PCF of all HTH lattices.   
The low density ($\rho = 0.06$) lattices provide a modest improvement in MCF and SPD, but also exhibit a much higher peak force than a HEX lattice. In particular, HTH lattice \#1 with CB as the dominant mechanism has an undesirably high peak force. 
This is because the microscale members participating in the CB mode are more slender ($\lambda=16$).  Consequently, the `$t$' and `$c$' members are farther apart, resulting in a larger moment arm, and therefore a higher peak force $\force_{p}$. 

\noindent
It should be noted that the final strain $\strain_{f}$ in these simulations is less than the densification strain (estimated to be $0.8\left(1-\rho\right)$ in~\cite{qiao_2016}). Therefore, the  trends in the SPD and MCF values are representative of the initial plateau region only. The present work, of course, focuses on the NB and CB modes, which manifest at strains of 4--8\%. Exploration of the high-strain (densification) response is currently underway. 
\subsection{The NB mode is insensitive to specimen size effects and symmetry boundary conditions}
\label{subsec::robustness}
\begin{figure}[h]
	\centering
	\begin{subfigure}{0.6\textwidth}
		\includegraphics[width=\textwidth]{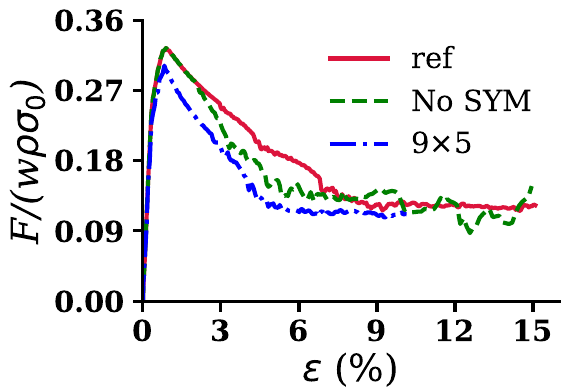}
	\end{subfigure}
	\caption{Robustness of the global $\force-\strain$ response. These plots show the effect of relaxing the symmetry boundary conditions (`No SYM') and using a lattice with a larger number of cells ($9\times 5$). The plot labelled `ref' corresponds to the reference simulation which uses symmetry and a lattice size of $5 \times 3$.}
	\label{fig::size_effect}
\end{figure}

The necking-buckling (NB) mode is insensitive to the use of half-symmetry or the size of the specimen in the FE simulations; in other contexts in cellular solids~\cite{onck_gibson_2001}, the number of unit cells / specimen size is known to affect the global response. 
In order to investigate such effects, a $5\times3$ HTH lattice with $\rho = 0.1, r = 7$ and symmetry boundary conditions is taken as a baseline simulation, and its $\force-\strain$ response studied (`ref' in Fig.~\ref{fig::size_effect}). Two other simulations, one with a larger lattice of $9\times 5$ cells, and a lattice without half-symmetry (`No SYM' in Fig.~\ref{fig::size_effect}) were conducted and their $\force-\strain$ responses plotted. 

\noindent
The global $\force-\strain$ responses are qualitatively similar, while still reflecting a size and symmetry effect on the global modes of deformation as expected. In particular, the global mode is not symmetric when the symmetry boundary condition is relaxed, and the deformation is more localized in the bigger lattice. 
Importantly, however, the dominant microscale mechanism in all three cases is the NB mode with no noticeable micro-mechanical changes, demonstrating the insensitivity of the NB and other local modes to the global deformation mode. This is also corroborated by the theoretical analysis of Sec.~\ref{sec:theory} which predicts that the NB mode can occur in a macroscale member regardless of the global deformation mode.

\section{Conclusion}
Continuum FE modelling, in tandem with a custom-built remeshing framework, enables the simulation of large plastic deformation of hierarchical lattices in the otherwise inaccessible low-moderate--slenderness regime. Remeshing is especially critical to accurately capture complex microscale deformation patterns ranging from self-contact and cusp formation to necking and ductile failure. 
The high-fidelity simulations revealed a new, local mode of failure which involves buckling of a microscale member accompanied by the necking and subsequent ductile failure of another microscale member. 
This necking-buckling (NB) mode is caused by truss-action, and NB itself is the primary driver of rotation of macro-cell walls in HTH lattices to accommodate global compressive strain.
The occurrence of the NB mode is insensitive to the global deformation modes and material fracture strain, and is also seen in an alternative, diamond-triangle hierarchical (DTH) lattice.
However, when the microscale members are very slender, necking-free coordinated buckling (CB) of a group of microscale members is seen to be preferred over the NB mode. 
A theoretical analysis of the force distribution, elastic buckling, and post-yield stability of an HTH macro-cell wall explain the occurrence of the NB and CB modes, and also establishes the range of relative density and scale ratios in which these modes are likely.
Activation of the NB mode in higher-density HTH lattices is seen to lead to generally better  crashworthiness. This has implications for the design of two-scale hierarchical lattices for such applications. 

\appendix
\renewcommand\thesection{Appendix \Alph{section}}
\section{Material properties}
\label{app::material}
\renewcommand\thesection{\Alph{section}}
The rate-hardening factor $\beta$ in Eq.~\eqref{eq::flow} is given by~\cite{pare_2016}
\begin{equation}
	\beta(\dot{\bar{\epsilon}}) = \left\{
	\begin{matrix}
		1& 0 \le \dot{\bar{\epsilon}} \le \dot{\bar{\epsilon}}_{1}\\
		1 + C_{1}\log(\dot{\bar{\epsilon}}/\dot{\bar{\epsilon}}_{1})  &\dot{\bar{\epsilon}}_{1} \le \dot{\bar{\epsilon}} < \dot{\bar{\epsilon}}_{2}\\
		\left(1 + C_{1}\log(\dot{\bar{\epsilon}}_{2}/\dot{\bar{\epsilon}}_{1})\right)\left(1 + C_{2}\log\left(\dot{\bar{\epsilon}}/\dot{\bar{\epsilon}}_{2}\right)\right) & \dot{\bar{\epsilon}}_{2} \le \dot{\bar{\epsilon}} < 10^5 \text{ s}^{-1}
	\end{matrix}		
	\right.
	\label{eq::rate}
\end{equation}
Table~\ref{tab::plastic} lists the values of the parameters appearing in Eqs.~\eqref{eq::flow} and~\eqref{eq::rate}. 
\begin{table}[h!]
	\centering	
	\caption{Plastic properties of AA5052-H32~\cite{lloyd_1982}}
	\begin{tabular}{|c |c| c| c| c| c| c| c |c|}
		\hline
		$\sigma_s$ (MPa) & $\sigma_0$  (MPa) & $N$ & $\bar{\epsilon}_{0}$ & $p$ & $C_1$ & $C_2$ & $\dot{\bar{\epsilon}}_{1}$ (s$^{-1}$) & $\dot{\bar{\epsilon}}_{2}$ (s$^{-1}$)\\
		\hline
		416.0 & 84.0 & 0.801 & 0.15 & 0.440 & 0.025 & 0.135 & 1.0 & 2400.0\\
		\hline
	\end{tabular}
	\label{tab::plastic}
\end{table}

\noindent
Table~\ref{tab::damage} presents the dependence of $\bar{\epsilon}_f$ on $\eta$ considered in this study. Failure under compression ($\eta < 0$) is prevented by choosing large values of $\bar{\epsilon}_f$. Studies~\cite{guo_2013} have shown that this material undergoes ductile failure under pure shear $\eta=0$ as well. The failure under tension ($\eta > 0$) is preceded by rapid necking which results in large, localized plastic strains. 
\begin{table}[h]
	\centering
	\caption{Assumed strain to failure $\bar{\epsilon}_f$ as a function of stress triaxiality $\eta$}
	\begin{tabular}{| c| c| c| c |}
		\hline
		$\eta$ & -0.3 & 0.0 & 0.3\\
		\hline
		$\bar{\epsilon}_f$ & 10.0 & 2.0 & 2.0\\
		\hline
	\end{tabular}
	\label{tab::damage}
\end{table} 

\renewcommand\thesection{Appendix \Alph{section}}
\section{Simulation quality}
\label{app::quality}
\renewcommand{\thesection}{\Alph{section}}
The evolution of FE energies (including KE, as mentioned in Sec.~\ref{subsec:bcs}) is useful in estimating the overall simulation quality. Table~\ref{tab::energies} presents the FE energies of a representative $r = 7, \rho = 0.10$ HTH lattice simulation at increasing strain levels. The energies are normalized with respect to the total external work done. Throughout the simulation, the artificial energy (AE) is maintained at a small fraction ($<10^{-2}$) of the net mechanical work done. AE is associated with hourglass control~\cite{belytschko_1991} of the reduced-integrated quadrilateral (CPE4R) elements in Abaqus, and its small value signifies that the mesh is well-refined. Further, the kinetic energy (KE) also remains at less than $10^{-3}$ of the work done at any point of the simulation, which indicates the absence of inertial effects despite mass-scaling~\cite{chung_1998}. Lastly, most of the external work done is dissipated by plasticity (PD) with a small fraction stored as elastic strain energy; this is characteristic of lattices in the range of densities examined in the present work.
\begin{table}[h!]
	\centering
	\caption{Normalized FE energies$^{**}$ of a typical ($r=7, \rho=0.10$) HTH lattice simulation; the AE and KE are very small throughout.}
	\begin{tabular}{|c|c|c|c|c|}
		\midrule[1 pt]
		$\strain$ (\%) & WK  & PD & AE & KE \\
		\midrule[1 pt]
		1 & 0.081 & 0.048 & 0.0001 & 0.0001  \\
		\hline
		4 & 0.421 & 0.391 & 0.0016 & 0.0002   \\
		\hline
		8 & 0.696 & 0.668 & 0.0026 & 0.0002   \\
		\hline
		\textbf{15} & \textbf{1.000} & \textbf{0.969} & \textbf{0.0036} & \textbf{0.0004}   \\
		\midrule[1 pt]
	\end{tabular}\\
	\footnotesize{$^{**}$WK: external work,  PD: plastic dissipation, AE: artificial energy, KE: kinetic energy}
	\label{tab::energies}
\end{table}

\subsection*{Competing interests}
We declare we have no competing interests.
\subsection*{Funding statement}
We acknowledge partial support of this work by grants from the High-Speed Rail Innovation Center (HSRIC) Trust and the ISRO-IISc Space Technology Cell, grant \# ISTC/CCE/NKS/489 to N.S.\hspace{0.15cm} N.C. was supported by a Prime Minister's Research Fellowship (PMRF) from the Ministry of Education, Government of India.

\label{sec:refs}
\bibliography{references}

\end{document}